\documentclass[11pt]{article}
\usepackage{mathtools,amsmath,mathrsfs}
\usepackage{booktabs}
\usepackage{amssymb}
\usepackage{cleveref}
\usepackage{subcaption}
\usepackage{natbib}

\usepackage[margin=1in]{geometry}

\usepackage{color}
\usepackage{ulem}
\usepackage{subcaption} 
\newcommand\RR{\mathbb{R}}

\DeclareMathOperator\trace{tr}
\DeclareMathOperator\cov{Cov}
\newcommand{\imgcell}[3][3.2cm]{%
  \begin{minipage}[t]{0.32\linewidth} 
    \centering
    \includegraphics[width=\linewidth,height=#1,keepaspectratio]{#2}\\[-0.3em]
    {\small #3}
  \end{minipage}%
}

\usepackage{siunitx}
\usepackage{endnotes}
\usepackage{tikz}
\usetikzlibrary{positioning}

\title{Contrastive Dimension Reduction: A Systematic Review}

\author{Sam Hawke$^{1,*}$, Eric Zhang$^{2,*}$, Jiawen Chen$^{3,4,*}$, Didong Li$^{2}$\\
Department of Mathematics and Statistics, Skidmore College$^1$ \\
Department of Biostatistics, University of North Carolina at Chapel Hill$^2$ \\
Gladstone Institutes$^3$\\
Department of Biomedical Data Science, Stanford University$^4$
}

\date{}
\begin{document}

\maketitle
\begingroup
\renewcommand\thefootnote{}\footnotetext{* These authors contributed equally to this work.}
\endgroup

\begin{abstract}
Contrastive dimension reduction (CDR) methods aim to extract signal unique to or enriched in a treatment (foreground) group relative to a control (background) group. This setting arises in many scientific domains, such as genomics, imaging, and time series analysis, where traditional dimension reduction techniques such as Principal Component Analysis (PCA) may fail to isolate the signal of interest. In this review, we provide a systematic overview of existing CDR methods. We propose a pipeline for analyzing case-control studies together with a taxonomy of CDR methods based on their assumptions, objectives, and mathematical formulations, unifying disparate approaches under a shared conceptual framework. We highlight key applications and challenges in existing CDR methods, and identify open questions and future directions. By providing a clear framework for CDR and its applications, we aim to facilitate broader adoption and motivate further developments in this emerging field.

\end{abstract}

\section{Introduction}

High-dimensional datasets are pervasive in modern data analysis across scientific disciplines, including genetics and genomics~\citep{bhola2018gene}, computer vision~\citep{shorten2019survey}, and wearable health monitoring~\citep{cho2021identifying, banaee2013data}. Such high dimensionality poses significant challenges, including high noise levels, computational inefficiency, redundant or correlated features, risk of overfitting, and the curse of dimensionality. Fortunately, many high-dimensional datasets are believed to concentrate near low-dimensional manifolds, a premise known as the manifold hypothesis, which is widely accepted and supported by empirical evidence~\citep{fefferman2016testing}. This observation motivates the use of dimension reduction (DR), which plays a central role in identifying low-dimensional structure embedded in high-dimensional data. DR improves signal-to-noise ratio~\citep{thudumu2020comprehensive}, enhances visualization and interpretability~\citep{johnstone2009statistical}, and reduces the computational cost of downstream tasks~\citep{fan2006statistical, fan2014challenges}. 

Driven by the growing scale and complexity of modern datasets, DR has received sustained attention over the past few decades, leading to a wide range of methods developed across applied mathematics, statistics, computational biology, machine learning, and various applied domains. Although these methods vary in their motivations, mathematical formulations, assumptions, and intended applications, they share the common goal of capturing meaningful low-dimensional structure. These developments have led to a rich landscape of DR methods, ranging from classical linear techniques to modern nonlinear and deep learning–based approaches. Classical linear methods such as Principal Component Analysis (PCA~\cite{hotelling1933analysis}) and multidimensional scaling (MDS~\cite{torgerson1952multidimensional}) aim to preserve global structure through projections or distance-preserving embeddings. To capture nonlinear structure, spectral methods such as Isomap~\citep{tenenbaum2000global}, Laplacian eigenmaps~\citep{belkin2003laplacian}, and diffusion maps~\citep{coifman2006diffusion} leverage graph-based representations to uncover manifold geometry. More recent algorithms such as t-SNE~\citep{van2008visualizing} and UMAP~\citep{mcinnes2018umap} prioritize local neighborhood preservation and are widely used for data visualization. Beyond these geometric approaches, deep learning has introduced autoencoders (AE~\cite{hinton2006reducing}) and variational autoencoders (VAE~\cite{kingma2014auto}), which learn nonlinear embeddings through neural networks and have proven effective in large-scale applications.

While traditional DR methods are effective for uncovering global or local structure within a single dataset, many scientific studies, particularly in biomedical research, are designed around a case-control framework. In such settings, the primary goal is not merely to capture dominant variation, but rather to identify structure that is unique to or enriched in one group (case, treatment, or foreground) relative to another (control, background). This contrastive objective arises naturally in a wide range of applications, yet standard DR methods are not tailored to isolate group-specific signals. This gap has motivated the development of contrastive dimension reduction (CDR) methods, where ‘contrastive’ refers specifically to distinguishing between case and control groups. 

\begin{figure}[!h]
  \centering
  \begin{subfigure}[t]{0.45\textwidth}
    \centering
    \includegraphics[width=\textwidth]{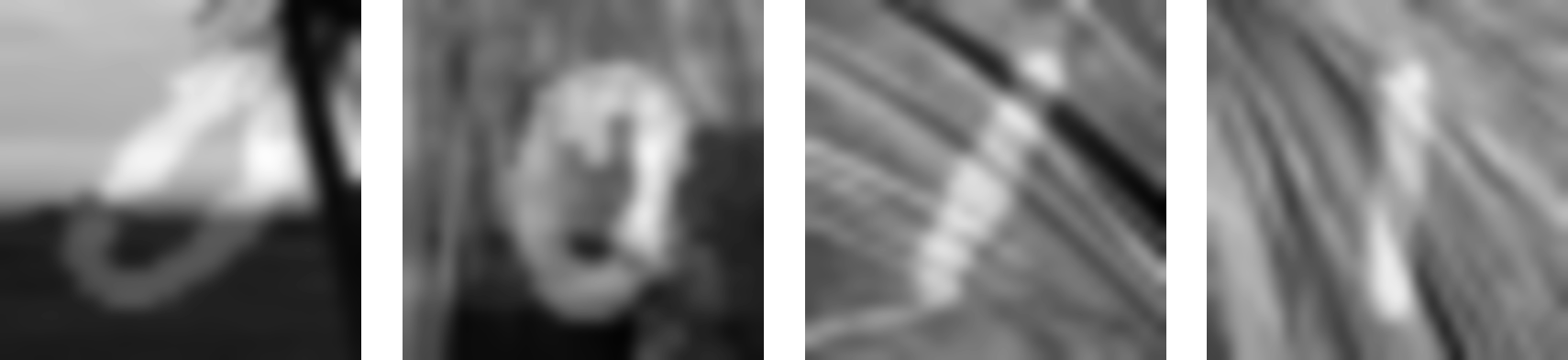}
    \vspace{0.001em}
    \caption{Foreground Data}
  \end{subfigure}
  \hfill
  \begin{subfigure}[t]{0.45\textwidth}
    \centering
    \includegraphics[width=\textwidth]{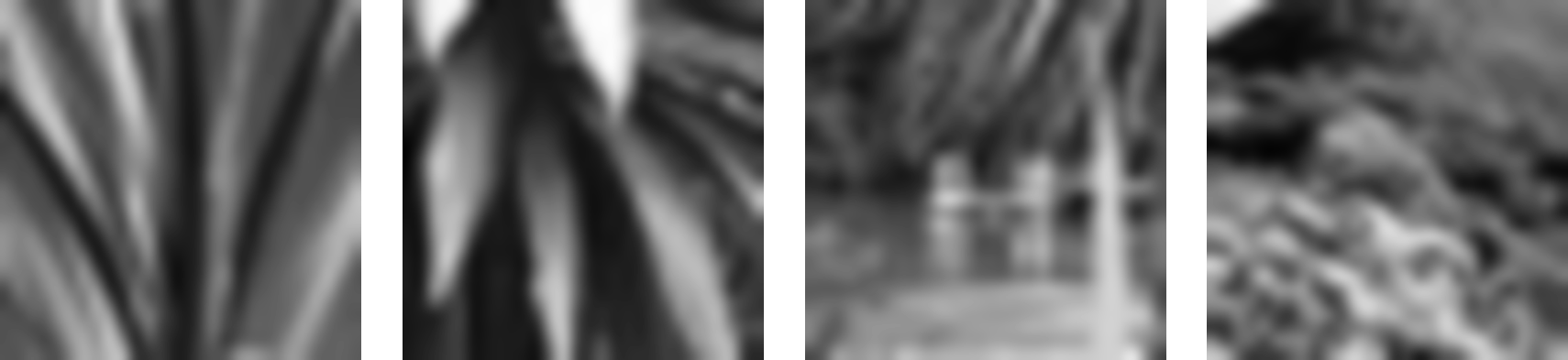}
    \vspace{0.001em}
    \caption{Background Data}
  \end{subfigure}

  \caption{Corrupted MNIST dataset. (a) Foreground data: MNIST digits 0 and 1 overlayed with grass images. (b) Background data: grass images.}
  \label{fig:mnist_ex}
\end{figure}

A representative toy example, widely used in the CDR literature, is the corrupted MNIST dataset~\citep{abid2018exploring}. As shown in Figure~\ref{fig:mnist_ex}, the foreground dataset is constructed by overlaying MNIST digits (0 and 1) onto natural background grass textures, while the background dataset consists solely of grass images. This design creates a structured foreground signal (the digits) embedded in high-variance background noise (the texture). The goal of CDR in this setting is to extract structure unique to the digit-containing images by leveraging the background dataset to remove shared texture variation. This illustrates the core idea of CDR: isolating meaningful signal that is unique to or enriched in the foreground while filtering out variation shared across both groups.

Motivated by such settings, a growing body of CDR methods has emerged in recent years.  Proposed approaches span a range of modeling paradigms, including extensions of classical linear methods, probabilistic formulations that incorporate uncertainty, and deep learning models designed to capture nonlinear contrastive features. While unified in their overarching goal, these methods differ substantially in their assumptions, algorithmic strategies, and applicability across domains.

Given the rapid growth of this emerging field, a systematic review of CDR methods is both timely and necessary. In this paper, we present a focused synthesis of existing CDR approaches, organized around several key contributions. First, we provide a systematic review of existing CDR methods. Second, we introduce a taxonomy (\Cref{fig:1}) that categorizes these methods based on their modeling assumptions and contrastive objectives, offering a unifying perspective on their relationships. Third, we aim to guide practitioners, particularly domain scientists, in selecting appropriate CDR tools for their specific research settings. Fourth, we illustrate how these methods work in practice using a toy dataset (corrupted MNIST) and a real-world dataset (mouse protein expression). Finally, we highlight current methodological limitations and identify open research questions that point to promising directions for future development in this area.

\section{Overview of CDR Methods}

In this section, we introduce notation and review key methods for CDR. We denote the foreground dataset as $X = \{x_1, \dots, x_{n_x}\} \subset \mathbb{R}^p$ and the background dataset as $Y = \{y_1, \dots, y_{n_y}\} \subset \mathbb{R}^p$ unless otherwise defined, where both datasets share the same ambient dimension $p$ but may differ in sample size and are not assumed to be paired. For simplicity, we assume each dataset has been centered independently. The methods we review differ in how they define and extract structure unique to the foreground data, and we organize them into linear and nonlinear categories, as well as CDR for data with additional structures, followed by a discussion of key preprocessing strategies that are orthogonal to these methods (\cref{fig:1}). We conclude this section with a table summarizing these methods and their characteristics, along with a taxonomic figure that illustrates their relationships and provides a practical pipeline for selecting and applying CDR methods (\cref{tab:sum}).

\subsection{Linear CDR Methods}\label{subsection:linear}

We first review linear CDR methods, which are often conceptually simpler, computationally efficient, and yield more interpretable low-dimensional representations compared to their nonlinear counterparts. Linear CDR methods seek to find a linear projection of the data $X \in \mathbb{R}^{n_x \times p}$ into a lower-dimensional subspace that emphasizes contrastive structure, i.e., signals unique to or enriched in the foreground dataset. These methods learn a loading matrix $V \in \mathbb{R}^{p \times d}$, yielding a reduced representation $XV\in\RR^{n_x\times d}$, where $d \ll p$ is the reduced dimension. In these cases, $V$ is generally constrained to lie on the Stiefel manifold, the space of all $p\times d$ matrices with orthonormal columns: $\text{St}(p, d) \coloneqq \{V \in \mathbb{R}^{p \times d} \mid V^\top V = I_d\}$, to ensure orthogonality of the projected directions. Linear CDR methods differ in their mathematical formulations, but they share this common algebraic structure. We further divide these linear methods into two subcategories: matrix decomposition–based methods and model-based methods.

\subsubsection{Matrix Decomposition--Based Methods}

Matrix decomposition--based methods form the core of linear CDR. They seek directions in which the foreground varies more than the background by modifying second--moment (covariance) information from the two groups. In most cases, the low-dimensional projection $V\in \text{St}(p, d)$ is obtained by solving a (generalized) eigen-problem. 
Variants may use low-rank matrix factorizations (e.g., singular value decomposition or CUR), but the shared idea is simple: adjust the second moments to highlight contrast between foreground and background. 

\subparagraph{Contrastive PCA (CPCA)}

CPCA~\citep{abid2018exploring} aims to uncover low-dimensional structure that is unique to or enriched in the foreground dataset $X$ relative to the background dataset $Y$. In the one-dimensional case, CPCA seeks a unit vector $v \in \mathbb{R}^p$ that maximizes variance in $X$ while penalizing variance in $Y$. Let $C_X = \frac{1}{n_x} \sum_{i=1}^{n_x} x_i x_i^\top$ and $C_Y = \frac{1}{n_y} \sum_{j=1}^{n_y} y_j y_j^\top$ denote the sample covariance matrices of the foreground and background datasets, respectively. CPCA solves the following optimization problem:
\begin{align*}
    \underset{\|v\| = 1}{\max}~v^\top C_X v - \gamma v^\top C_Y v \eqqcolon \underset{\|v\| = 1}{\max}~v^\top C v,
\end{align*}
where $\gamma \in [0, \infty]$ is a tuning parameter, and $C = C_X - \gamma C_Y$ is known as the contrastive covariance matrix. The solution is the leading eigenvector of $C$, i.e., the eigenvector corresponding to the largest eigenvalue. Notably, when $\gamma = 0$, CPCA reduces to PCA on $X$, and as $\gamma \to \infty$, the method recovers the direction of minimal background variance, i.e., orthogonal to the PCA results on $Y$.

In the multi-dimensional setting, CPCA seeks a matrix $V\in\text{St}(p, d)$ that maximizes the explained variance in $X$ while penalizing the explained variance in $Y$, by solving the following optimization problem:
\begin{align*}
    \underset{V \in \text{St}(p, d)}{\max}~\operatorname{tr}(V^\top C V).
\end{align*} The solution consists of the top $d$ eigenvectors of $C$ corresponding to the largest $d$ eigenvalues, analogous to standard PCA. Throughout this article, we assume eigenvalues are sorted in descending order.

\subparagraph{Generalized Contrastive PCA (GCPCA)}
GCPCA~\citep{de2024identifying} was introduced to address key limitations of CPCA, namely its dependence on a manually tuned contrastive parameter $\gamma$ and its asymmetric treatment of foreground and background datasets. To penalize high-variance dimensions and thereby remove the need for hyperparameter tuning, GCPCA solves the following objective function:
\begin{align*}
    \underset{V\in\mathrm{St}(p,d)}{\max} \quad \frac{\trace( V^\top(C_X - C_Y) V)}{\trace (V^\top (C_X + C_Y) V)},
\end{align*}
which is equivalent to computing the leading eigenvectors of $M^{-1}(C_X - C_Y)M^{-1}$, where $M = (C_X + C_Y)^{1/2}$. The resulting $V$ maximizes relative variance differences between $X$ and $Y$ in a fully symmetric fashion. This framework maximizes relative, rather than absolute, variance differences between $X$ and $Y$. The GCPCA framework also has the following useful variants:  
\begin{itemize}
    \item \textbf{GCPCA v2:} Maximizes the variance ratio $\frac{\trace(V^\top C_X V)}{\trace(V^\top C_Y V)}$, which is analogous to the generalized eigenvalue formulation used in classical methods such as Fisher’s linear discriminant analysis~\cite{zhao2024linear}. It identifies directions where the foreground dataset $X$ exhibits large variance relative to the background dataset $Y$, but in a multiplicative manner rather than additive as CPCA. 
    \item \textbf{GCPCA v3:} Maximizes the relative change $\frac{\trace(V^\top (C_X - C_Y) V)}{\trace(V^\top C_Y V)}$, which measures the relative increase in variance of $X$ compared to $Y$, normalized by the background variance. This variant is especially useful when the goal is to highlight dimensions where changes in variability are best interpreted relative to the baseline or control group.
\end{itemize}

These formulations provide a flexible, hyperparameter-free approach for CDR.

\subparagraph{Contrastive CUR (CCUR)}
Although linear DR methods offer a degree of interpretability, their loadings, i.e., the columns of $V$, represent linear combinations of all features, which can obscure direct interpretation in certain applications. To overcome this limitation, the CUR decomposition provides an interpretable alternative by selecting actual rows and columns from the data matrix~\cite{mahoney2009cur}. 

CUR decomposition approximates a matrix $X \in \mathbb{R}^{n_x \times p}$ by selecting representative columns $C$ and rows $R$, such that $X \approx CUR$, where $U$ is usually a dense matrix obtained by minimizing $\|X-CUR\|$ with respect to $U$. Columns and rows are typically selected based on leverage scores, which measure the importance of each column or row in the low-rank structure of $X$. This formulation offers interpretable approximations while simultaneously selecting representative rows and columns, a capability absent in traditional PCA.

CCUR~\citep{zhang2025contrastive1} extends this framework to a contrastive setting, where the goal is to identify columns and rows that are uniquely important to a foreground group relative to a background group. CCUR first computes leverage scores for both groups. Specifically, the leverage score for column \(j\) in each group is:
\[
l^X_j = \sum_{k=1}^K (v_j^{X,k})^2, \quad l^Y_j = \sum_{k=1}^K (v_j^{Y,k})^2,
\]
where \(v_j^{X,k}\) and \(v_j^{Y,k}\) are entries of the $j$-th right singular vectors of \(X\) and \(Y\), respectively, and \(K\) is the number of singular vectors retained. A contrastive score is then computed as:
\[
s_j = \frac{l^X_j}{l^Y_j + \epsilon},
\]
where \(\epsilon > 0\) is a small constant for numerical stability. Columns with the highest $d$ contrastive scores are selected as they exhibit strong influence in the foreground while having minimal impact in the background. Rows are selected by running CUR on the subset of columns selected by the aforementioned method and returning $R$. As a result, CCUR identifies features and samples that are most salient to the foreground group, highlighting patterns that distinguish it from the background.


\subsubsection{Model-Based Methods}

Although deterministic CDR methods can offer useful insights, they often fall short when datasets are noisy, incomplete, or when uncertainty about the embedding is important. In such settings, probabilistic models provide a principled framework that not only accommodates noise and missingness but also yields uncertainty quantification. This section introduces three representative model-based CDR methods that illustrate these advantages.

\subparagraph{Spectral Methods}
While not a CDR method per se, \citep{zou2013contrastive} presents an early probabilistic framework with a contrastive mechanism that inspired later CDR approaches such as CPCA and CLVM. Their model assumes a mixture distribution of the form\[
    p(x) \;=\; \sum_{j=1}^J w_j f(x;\theta_j),
\]
where $J$ is the number of mixture components, $f(\cdot;\theta_j)$ is the
density of component $j$ with parameters $\theta_j$, and $w_j$ is its mixing
weight. Suppose the foreground distribution is generated by components
indexed by $A \cup B$, and the background by $B \cup C$, with $A,B,C\subset \{1,\cdots,J\}$
disjoint index sets. The contrastive goal is to recover the
foreground-specific components indexed by $A$, without explicitly learning a
model for the background.

This method is based on method-of-moments, which identifies mixture
components via empirical second- and third-order moments. Let
$M^{(f)}_2, M^{(f)}_3$ denote the foreground moments and
$M^{(b)}_2, M^{(b)}_3$ the background moments. For a contrastive parameter
$\gamma > 0$, the modified moments are
\[
    M_2 = M^{(f)}_2 - \gamma M^{(b)}_2, 
    \qquad
    M_3 = M^{(f)}_3 - \gamma M^{(b)}_3,
\]
where $\gamma$ controls the strength of background suppression: $\gamma=0$
reduces to the standard spectral decomposition of the foreground moments,
while larger $\gamma$ highlights features distinctive to the foreground. 

Although this method is not framed as a dimension reduction tool, its core idea, subtracting background moments to isolate foreground structure, serves as an early prototype for later CDR methods.

\subparagraph{Probabilistic Contrastive PCA (PCPCA)}

PCPCA~\citep{li2020probabilistic} places CPCA in a probabilistic latent variable framework, enabling robustness to noise and missingness, and principled inference for uncertainty quantification. The model assumes both foreground and background arise from a shared low-dimensional linear latent space:
\[
x = W z_x + \varepsilon_x, \qquad y = W z_y + \varepsilon_y,
\]
with \(z_x,z_y \sim \mathcal{N}(0,I_d)\), loading matrix \(W \in \mathbb{R}^{p\times d}\), and Gaussian noise \(\varepsilon_x,\varepsilon_y \sim \mathcal{N}(0,\sigma^2 I_p)\).

Rather than maximizing a single likelihood, PCPCA fits \(W\) and \(\sigma^2\) by maximizing a contrastive likelihood that favors models which explain the foreground well and the background poorly:
\[
\arg\max_{W,\sigma^2} \ \frac{p(X \mid W,\sigma^2)}{p(Y \mid W,\sigma^2)^{\gamma}}
\;=\; \arg\max_{W,\sigma^2} \Big\{ \sum_{x\in X}\log p(x\mid W,\sigma^2) \;-\; \gamma \sum_{y\in Y}\log p(y\mid W,\sigma^2) \Big\}.
\]
This objective can be read as foreground log-likelihood minus a \(\gamma\)-weighted background log-likelihood. 

Under the Gaussian model, optimizing the objective above yields closed-form estimators in terms of the eigenpairs of the contrastive covariance \(C = C_X - \gamma C_Y\). Let $(V,\Lambda=\mathrm{diag}(\lambda_1,\cdots,\lambda_d))$ be the eigenpairs of $C$, then the solution of PCPCA is given by 
\[\widehat{W}=V\left(\frac{\Lambda}{n_x-\gamma n_y}-\widehat{\sigma}^2\mathrm{Id}_d\right)^{1/2},~~\widehat{\sigma}^2=\frac{1}{(n_x-\gamma n_y)(p-d)}\sum_{j=d+1}^p \lambda_j.\]

As special cases, when \(\gamma=0\), PCPCA becomes Probabilistic PCA (PPCA~\cite{tipping1999probabilistic}) on \(X\); when \(\sigma^2 \to 0\), PCPCA recovers CPCA.
The probabilistic formulation also supports generalized Bayesian inference via a Gibbs posterior over \(W,\sigma^2\) and can accommodate missing entries via gradient-based optimization, while keeping the same contrastive second-moment intuition as CPCA.

\subparagraph{Contrastive Latent Variable Model (CLVM)}

CLVM~\citep{severson2019unsupervised} proposes a probabilistic latent variable model:
\[
x_i = S z_i + W t_i + \mu_x + \varepsilon_i, \quad y_j = S z_j + \mu_y + \varepsilon_j,
\]
where $z_i, z_j \in \mathbb{R}^k$ are latent variables capturing structure shared across groups, $t_i \in \mathbb{R}^d$ are latent variables unique to the foreground, and $S \in \mathbb{R}^{p \times k}$ and $W \in \mathbb{R}^{p \times d}$ are corresponding factor loading matrices. The residuals $\varepsilon_i, \varepsilon_j$ are assumed to be Gaussian noise terms. Under this model, the marginal covariance of the foreground data is $SS^\top + WW^\top + \sigma^2 I$, while the background data has marginal covariance $SS^\top + \sigma^2 I$, allowing $W$ to capture variation specific to the foreground group.

Parameter estimation can be performed using expectation-maximization (EM) under Gaussian assumptions or via variational inference (VI) for more general likelihoods and priors. The model admits several useful extensions, including sparse CLVM for automatic feature selection, model selection via automatic relevance determination (ARD) priors on $S$, and robust CLVM using Student-$t$ likelihoods to handle outliers. Across these formulations, the primary goal remains the same: to recover a low-dimensional latent representation of the foreground-specific structure while accounting for shared structure.

\subparagraph{Contrastive Poisson Latent Variable Model (CPLVM)}
In the above linear models, Gaussian distributions are commonly assumed, which may be inappropriate for genomic data (e.g., gene expression) where observations are nonnegative counts. To address this gap, CPLVM~\citep{jones2022contrastive} extends CLVM to model count-based data by using a Poisson likelihood instead of a Gaussian. To account for differences in sequencing depth, cell-specific size factors $\alpha_i^b$ and $\alpha_j^f$ are introduced for each background and foreground cell, respectively. In addition, gene-specific multiplicative scale parameters $\delta \in \mathbb{R}_{+}^p$ model mean shifts in expression between conditions. The generative model is defined as
\[
y_i \mid z_i^{b} \sim \operatorname{Poisson}\!\left(\alpha_i^b\, \delta \odot (S^\top z_i^{b})\right), 
\qquad
x_j \mid z_j^{f}, t_j \sim \operatorname{Poisson}\!\left(\alpha_j^f\, (S^\top z_j^{f} + W^\top t_j)\right),
\]
where $\odot$ is the Hadamard product, $z_i^{b}, z_j^{f} \in \mathbb{R}_{+}^{k}$ capture shared structure, $t_j \in \mathbb{R}_{+}^{d}$ captures foreground-specific structure, and $S, W$ are corresponding loading matrices. Gamma priors are placed on the latent variables and loadings, while $\delta$ follows a log-normal prior. Inference is performed using stochastic variational inference with mean-field log-normal variational distributions. By directly modeling raw counts and incorporating $\delta$, CPLVM isolates structured changes in gene expression unique to the foreground condition while controlling for shared and technical sources of variation.

\subsection{Nonlinear CDR Methods}

Linear contrastive methods are effective when foreground structure is well approximated by a low-dimensional linear subspace of $\mathbb{R}^p$. They can struggle, however, when meaningful variation lies on a curved manifold~\citep{van2008visualizing, bengio2013representation, cunningham2014dimensionality}. Nonlinear CDR approaches extend the same core idea: highlight what is salient in the foreground and deemphasize what is shared with the background, using flexible function classes such as deep neural networks. 
The subsections below sketch these nonlinear approaches and how the contrastive mechanism is enforced.

\subparagraph{Contrastive Variational Autoencoder (CVAE)}

CVAE \citep{abid2019contrastive} casts CDR in a deep variational autoencoder. Each sample is described by a shared latent $z$ and a salient latent $s$. Foreground points use both codes $z$ and $s$, while background points use only the shared code $z$ while setting $s\equiv 0$:
\[
x \sim f_\theta(s,z), \qquad y \sim f_\theta(0,z),
\]
where $f_\theta$ is a neural decoder. Two encoders $q_{\phi_s}(s\mid x)$ and $q_{\phi_z}(z\mid \cdot)$ infer the latents; for background, only $q_{\phi_z}(z\mid y)$ is used. Training maximizes a sum of evidence lower bounds (ELBOs): a standard VAE ELBO on foreground (both $s$ and $z$) and an ELBO on background with $s$ fixed to zero. This encourages $s$ to carry foreground-specific information while $z$ captures structure shared across groups. The learned $s$ provides a nonlinear low-dimensional embedding of foreground-specific variation.

\subparagraph{Contrastive Variational Inference (CVI)}

CVI \citep{weinberger2023isolating} adopts the same shared/salient split, but specifically designed for gene expression data, with a likelihood appropriate for counts. Each sample has $z$ (shared) and $t$ (treatment-specific). Foreground observations depend on $(z,t)$, while background observations set $t=0$:
\[
x \sim f_\theta(z,t) \ \text{(foreground)}, \qquad y \sim f_\theta(z,0) \ \text{(background)},\qquad z,t \sim \mathcal{N}(0,I).
\]
Amortized variational inference learns encoders for $z$ and $t$; the shared decoder $f_\theta$ maps latents to distributional parameters (for example, negative binomial means for counts). By explicitly enforcing $t=0$ in the background term of the objective, CVI separates treatment-specific from shared effects and yields a nonlinear embedding in the $t$-space. 

\subparagraph{Contrastive Feature Selection (CFS)}
CFS \citep{weinberger2023feature} extends the 
contrastive feature selection framework (see also CCUR) to nonlinear settings, with the goal of identifying a 
small set of target features that capture residual variation specific to the 
salient signal after background variation has been explained. 

The method operates in two stages. In the first stage, a background 
encoder--decoder pair $(g,h)$ is trained solely on background samples, so that 
the encoder produces a low-dimensional representation 
$b = g(y;\phi) \in \mathbb{R}^d$ of nuisance variation:
\[
\min_{\phi,\eta}\ \mathbb{E}_{y}\,
  \big\|h(g(y;\phi);\eta)-y\big\|^2.
\]
This encoder is subsequently fixed and used to provide background summaries 
$b$ for the target data. 

The second stage addresses feature selection. We hope to identify a
subset $S \subseteq \{1,\cdots,p\}$ of features from $x \in \mathbb{R}^p$ that, together 
with $b$, best reconstructs $x$. Because direct optimization over discrete 
subsets is combinatorially hard for high dimensional data, CFS adopts a differentiable relaxation based on 
stochastic gates. Each feature $x_i$ is multiplied by a 
gate $G_i \in [0,1]$ defined as
\[
G_i = \max\!\big(0,\,\min(1,\ \mu_i + \zeta)\big),
\quad \zeta \sim \mathcal{N}(0,\sigma^2),
\]
where $\mu_i$ is a learnable mean. The gated features $x \odot G$ serve as a 
continuous surrogate for the discrete subset $S$. To encourage sparsity, a 
penalty on the expected number of active gates is included, leading to the 
optimization problem
\[
\min_{\theta,\,\mu}\ \mathbb{E}_{x}\,
  \big\|f_\theta(b,\ x \odot G)-x\big\|^2
  \;+\; \lambda \sum_{i=1}^p \Phi\!\left(\tfrac{\mu_i}{\sigma}\right),
\]
where $\odot$ denotes the Hadamard product, $\lambda$ controls the degree of 
sparsity, and $\Phi$ is the standard Gaussian CDF. 

This relaxation transforms feature selection into a differentiable procedure 
that can be trained end-to-end alongside the reconstruction network. As a nonlinear counterpart of CCUR, CFS identifies features that explain foreground-specific variation, offering improved interpretability.


\subsection{CDR for Data with Additional Structure}
In certain applications, the data possess additional structure beyond the standard setup with $X\in\mathbb{R}^{n_x\times p}$ and $Y\in\mathbb{R}^{n_y\times p}$. One common example is functional data, where each sample is a function rather than a finite-dimensional vector. Another example is supervised settings, where a response variable is available. These additional structures can be leveraged to guide CDR more effectively. In this subcategory, we present three representative methods: one that adapts CDR to functional data, and two that incorporate supervision when a response variable is available.


\subparagraph{Contrastive Functional PCA (CFPCA)}
CFPCA~\citep{zhang2025contrastive2} extends CPCA to the setting of functional data, where each observation is a real-valued function. Instead of working with finite-dimensional vectors, CFPCA seeks functional directions that distinguish the foreground group from the background group. We use functions over $\RR$ as an illustrative example, where functions in this case are curves. Let \(\{ x_i(t)\}_{i=1}^{n_x}\) and \( \{y_j(t)\}_{j=1}^{n_y}\) denote curves in the foreground and background groups, respectively, and let \( C_X(t,s) \) and \( C_Y(t,s) \) denote their sample covariance functions. CFPCA identifies a function \( v(t) \in L^2(\mathbb{R}) \) that maximizes the foreground variance while penalizing variance in the background:
\[
\underset{\|v\| = 1}{\arg\max} \int \int \left(C_X(t, s) - \gamma C_Y(t, s)\right) v(t)v(s) \, ds\, dt,
\]
where $\gamma \geq 0$ controls the degree of background suppression. When $\gamma = 0$, CFPCA reduces to standard FPCA. As $\gamma \to \infty$, the solution lies in the directions orthogonal to those with high background variance.
In practice, when curves are observed at a finite number of time points, these functions are represented as vectors $x_i(t_k),y_j(t_k)$, and the covariance operators become empirical covariance matrices $C_X$ and $C_Y$. If curves are aligned and observed on a common time grid, the integral operator can be approximated by matrix multiplication, and the associated eigenproblem becomes:
\[
w C v = \lambda v,
\]
where \( C = C_X - \alpha C_Y \) is the contrastive covariance matrix estimated from discretized data, and \( w \) is a constant related to the time grid spacing. The solution \( v \) can be interpreted as a discrete approximation to the contrastive eigenfunction, and optionally smoothed via interpolation.
CFPCA provides a natural extension of CPCA to time series, uncovering dynamic patterns enriched in the foreground group relative to background temporal variation.

\subparagraph{Contrastive Inverse Regression (CIR)}
In some applications, a response variable is available, giving rise to the supervised CDR setting. To understand this case, we first review supervised DR. A notable example is Sliced Inverse Regression (SIR~ \citep{li1991sliced}), which assumes the response $y$ depends on the covariates $x$ only through a low-dimensional projection $V^\top x$, i.e., $y = f(V^\top x) + \epsilon$, where $\epsilon$ is noise and $f$ is an arbitrary function. A key insight is that, under mild assumptions, the inverse regression curve $m(y) \coloneqq \mathbb{E}[X \mid Y = y]$ lies in the subspace spanned by $V$. Therefore, learning $V$ reduces to calculating eigenspace of $\mathrm{Cov}(m(y))$, which can be approximated via slicing $y$.

To extend this idea to the supervised CDR setting, CIR \citep{hawke2023contrastive} borrows the inverse-regression framework of SIR and adapts it to a contrastive setting. Since $y$ is reserved for the response, we now use $(X, y)$ for the foreground data and $(\widetilde X, \widetilde y)$ for the background data.
Let $C_X$ and $C_{\widetilde X}$ be the covariance matrices of two groups,
and let $m_y=\mathbb{E}[X\mid y]$, $\widetilde m_{\widetilde y}=\mathbb{E}[\widetilde X\mid \widetilde y]$ be the inverse regression curves.
CIR seeks a subspace that preserves the ability to predict the response variable $y$ in the
foreground while suppressing that in the background by minimizing
\[
L(V)
= \mathbb{E}_y\!\left[\big\|m_y - P_{C_XV} m_y\big\|^2\right]
  - \gamma\,\mathbb{E}_{\widetilde y}\!\left[\big\|\widetilde m_{\widetilde y} - P_{C_{\widetilde X}V}\widetilde m_{\widetilde y}\big\|^2\right],
\]
where \(P_{C V}\) is the projection onto \(\mathrm{span}(C V)\) and \(\gamma\ge0\) controls
the strength of background subtraction. This loss is equivalent to a
difference of SIR–type loss,
\[
L(V)= -\,\mathrm{tr}\!\left(V^\top A V\,\left(V^\top C_X^2 V\right)^{-1}\right)
      + \gamma\,\mathrm{tr}\!\left(V^\top \widetilde A V\,\left(V^\top C_{\widetilde X}^2 V\right)^{-1}\right),
\]
where \(A=C_X\cov(m_y)C_X,\widetilde A=C_{\widetilde X}\cov(\widetilde{m}_{\widetilde{y}})C_{\widetilde X}\), which can be estimated via slicing in practice.

When \(\gamma=0\) (no contrastive term) CIR reduces to SIR: reparameterizing
with \(W=C_XV\) orthonormalizes the columns and yields a standard
eigenproblem for \(\mathrm{Cov}\!\big(m(Y)\big)\).
For \(\gamma>0\), however, the objective involves both
\(V^\top C_X^2 V\) and \(V^\top C_{\widetilde X}^2 V\). In this case, no single change of variables can
simultaneously whiten \(C_X\) and \(C_{\widetilde X}\), so the closed-form reduction to
an eigenproblem is lost. CIR therefore relies on numerical algorithms to solve a constrained optimization on
the Stiefel manifold \(\mathrm{St}(p,d)\). 


\subparagraph{Contrastive Linear Regression (CLR)} 



Beyond CDR, many applications aim directly at predicting a response variable observed only in the foreground group. For instance, the foreground group may consist of treated subjects with observed treatment responses, while the background group lacks such responses due to the absence of treatment.

\begin{table}[h!]
\centering
\caption{CDR methods and their key characteristics}
\begin{tabular}{@{}llcccccc@{}}
\toprule
\textbf{Method} & \textbf{Year} & \textbf{Linear} & \textbf{Probabilistic} &  \textbf{Additional Structures} & \textbf{Feature Selection} \\
\midrule
CPCA & 2018 & \checkmark &  &  &   \\
CLVM & 2019 & \checkmark & \checkmark &     &    \\
CVAE & 2019 &   &  \checkmark &  &    \\
CPLVM & 2022 & \checkmark  & \checkmark &   & \\
CVI & 2023 &   & \checkmark  &  &  \\
CFS & 2023 &   &  &   & \checkmark  \\
PCPCA & 2024 & \checkmark &  \checkmark    &   &     \\
CIR & 2024 & \checkmark &  &  \checkmark    &    \\
GCPCA & 2024 &  \checkmark &   &   &          \\
CCUR & 2025 &  \checkmark &   &   & \checkmark        \\
CFPCA & 2025 & \checkmark &  &  \checkmark     &     \\

CLR & 2025 & \checkmark & & \checkmark &  \\

\bottomrule
\end{tabular}\label{tab:sum}
\end{table}

CLR~\citep{zhang2024contrastive} is specifically designed for this setting. It assumes that the response variable in the foreground is determined by a low-dimensional signal that is unique to the foreground group, consistent with the core principle of aforementioned CDR methods. Formally, let foreground observations $\{(x_i, r_i)\}_{i=1}^{n_x} \subset \mathbb{R}^p \times \mathbb{R}$, where $r$ is the response variable for subject $i$, and background observations $\{y_j\}_{j=1}^{n_y} \subset \mathbb{R}^p$ without observed responses. The CLR model is:
\begin{align*}
    x &= Sz_a + Wt + \epsilon_a, \quad 
    y = Sz_b + \epsilon_b, \quad 
    r = \beta^\top t + \eta,
\end{align*}
where $S,W \in \mathbb{R}^{p \times d}$ are loading matrices, $z_a, z_b, t \sim \mathcal{N}(0, \mathrm{I}_d)$ represent shared and foreground-specific latent factors, and $\epsilon_a, \epsilon_b, \eta$ are Gaussian noise terms. Here, $S$ captures shared structure between foreground and background, while $W$ captures the variation specific to the foreground. The regression coefficient $\beta$ links the salient foreground representation $t$ to the response $r$. Estimation proceeds by maximizing the likelihood over parameters $\theta = (S,W,\beta,\sigma^2,\tau^2)$.

Unlike CIR, which seeks contrastive subspaces for interpretability, CLR directly models the predictive relationship between foreground covariates and their response, after removing variation shared with the background. This formulation prioritizes foreground-specific associations and improves generalizability in high-dimensional prediction tasks.

To this end, we summarize the aforementioned CDR methods together with their key characteristics in \Cref{tab:sum}.

\subsection{Pre-Processing Steps}

While the CDR methods described above provide a diverse and powerful set of tools for extracting foreground-specific signals, several practical questions need to be addressed before implementing them. First, when multiple candidate background datasets are available, how should one define foreground versus background? Second, is there meaningful variation unique to the foreground group? If not, i.e., if the foreground and background share the same structure, then CDR may be unnecessary, and standard DR may suffice. Third, if foreground-specific structure exists, how should the reduced dimension $d$ be chosen? This tuning parameter appears across nearly all methods and governs the fidelity and interpretability of the representation. In this section, we discuss several efforts in this direction as preprocessing steps prior to applying CDR methods.


\subparagraph{Background Selection}
In certain applications, defining the background group is nontrivial due to the presence of multiple candidate datasets. A critical step in such settings is the selection of background datasets. 
A valid background should capture only the structure that is common with the foreground, without introducing additional dataset-specific variation that could confound contrastive inference. 

BasCoD \citep{park2025systematic} provides a principled statistical framework for evaluating candidate backgrounds.  
Denote the foreground dataset by
$X_0  \in \mathbb{R}^{n_0\times p},$
 and each candidate background dataset by
$X_j  \in \mathbb{R}^{n_j\times p}$ for $j \in C$, where $C$ is the index set of candidate backgrounds. For each $j \in \{0\}\cup C$, the model is 
\[
x_j = f_j(c_j,s_j,\epsilon_j),
\]
where $c_j \in \mathbb{R}^{d_c}$ are shared latent embeddings, 
$s_j \in \mathbb{R}^{d_{ j}}$ are dataset-specific embeddings, 
and $\epsilon_j \in \mathbb{R}^p$ is Gaussian noise. 
The foreground depends on both shared and specific components $(c_0,s_0)$, 
while a valid background depends only on the shared component. 
Thus, a valid background $X_j$ satisfies
\[
x_j = f_j(c_j,0,\epsilon_j), \quad j \in B,
\]
where the set of valid backgrounds is $B := \{\, j \in C : s_j=0 \,\}.$

In the linear setting, the model can be simplified as

\begin{align}
x_j &= \Gamma_c c_j + \Gamma_{s,j}s_j + \epsilon_j, \qquad j \in \{0\}\cup C \nonumber.
\end{align}
Let $\Gamma_j := [\Gamma_c,\,\Gamma_{s,j}] \in \mathbb{R}^{p\times (d_c+d_j)}$ and $P_0$ denote the orthogonal projection matrix onto $\mathcal{C}(\Gamma_0)$, 
the column space of the foreground loading matrix $\Gamma_0$. 
If $j \in B$, then $s_j=0$ and hence $x_j = \Gamma_c c_j + \epsilon_j$.  
This implies $\mathcal{C}(\Gamma_j) = \mathcal{C}(\Gamma_c) \subseteq \mathcal{C}(\Gamma_0)$
or equivalently $\Gamma_j = P_0 \Gamma_j.$
Therefore, for any candidate $j\in B$, the null hypothesis that $X_j$ is a valid background is

\
\[
H_{0,j}: \Gamma_j = P_0 \Gamma_j.
\]

To test this hypothesis, BasCoD computes sample correlations between each column of $\Gamma_j$ and its projection by $P_0$, stabilizes them using a Fisher transformation, and combines the results via Fisher’s method to yield a $\chi^2$ test statistic. Small $p$-values suggest $X_j$ contains variation not shared with the foreground and should be excluded as a background.

In nonlinear settings such as CVI or CVAE, the loading matrix \(\Gamma_j\) is not well-defined for general decoder $f_j$. To adapt the BasCoD procedure used in the linear case, a linear approximation to the nonlinear embedding is obtained by regressing the observed data \(X_j\) onto the low-dimensional latent representation \(L_j\) learned by the model:
\[
\widehat\Gamma_j \;=\; \arg\min_{B\in\mathbb{R}^{p\times d_j}} \; \|X_j - L_j B^\top\|_2^2,
\]
where \(L_j \in \mathbb{R}^{n_j\times d_j}\) denotes the latent embeddings for dataset \(j\).  

\subparagraph{Contrastive Dimension Estimation (CDE)}
After selecting an appropriate background dataset, the next question is to determine how many foreground-specific directions are unique to or enriched in the foreground relative to the (selected) background. CDE~\citep{hawke2024contrastive} separates this problem into two tasks: first, a hypothesis test for the existence of any contrastive structure; second, an estimator of its dimension when present. 

The problem is formulated as a linear latent variable model
\[
x_i=S_x z_i+\varepsilon_i,\qquad y_j=S_y w_j+\varepsilon_j,
\]
where \(S_x\in\mathbb{R}^{p\times d_x}\) and \(S_y\in\mathbb{R}^{p\times d_y}\) are full rank loading matrices, \(z_i\sim\mathcal N_{d_x}(0,I)\), \(w_j\sim\mathcal N_{d_y}(0,I)\), and \(\varepsilon_i,\varepsilon_j\) are Gaussian noise. Let \(V_x\) and \(V_y\) be the left singular matrices of $S_x$ and $S_y$, respectively, the contrastive subspace and contrastive dimension are defined as
\[
V_{xy}\;:=\;\mathrm{Proj}_{V_y^\perp}(V_x),\qquad d:=\dim(\mathcal{C}(V_{xy})).
\]
Then absence of unique information in foreground is:
\[\mathcal{C}(V_x)\subset \mathcal{C}(V_y)\Longleftrightarrow \mathcal{C}(V_{xy})=\{0\}\Longleftrightarrow d=0.\]

As a result, the hypothesis testing problem becomes  

\[H_0:\,d=0~~\text{vs}~~H_1:\,d>0.\]

Then define $\lambda_k $ be the $k$-th singular value of $V_x^\top V_y$ and $\theta_k\coloneqq \arccos(\lambda_k)$ is known as the principal angle, where $k=1,\cdots, \min(d_x,d_y)$. CDE constructs the test statistics via the maximal principal angles between \(V_x\) and \(V_y\), denoted by \(\theta_{\max}\coloneqq\max_{k=1,\cdots,\min(d_x,d_y)}\theta_k\), or equivalently, the smallest singular value $\lambda_{\min}\coloneqq \min_{k=1,\cdots,\min(d_x,d_y)}\lambda_k$. Larger $\theta_{\max}$ or smaller $\lambda_{\min}$ indicate greater distinction between $V_x$ and $V_y$. 
Significance is assessed by a contrastive bootstrap that enforces the null: for \(b=1,\dots,B\), resample a foreground \(X^{(b)}\) with replacement from \(X\) and resample a background \(Y^{(b)}\) with replacement from the pooled set \(X\cup Y\); compute \(\lambda_{\min}^{(b)}\) exactly as for the observed data. The p-value is
\[
 p\;\coloneqq\;\frac{1}{B}\sum_{b=1}^B \mathbf{1}\!\left\{\lambda_{\min}^{(b)}<\lambda_{\min}\right\},
\]
so small \(p\) indicates unusually small alignment (large angle) under \(H_0\), and we reject in favor of \(d>0\).


\begin{figure}[!h]
    \centering
    \includegraphics[width=\linewidth]{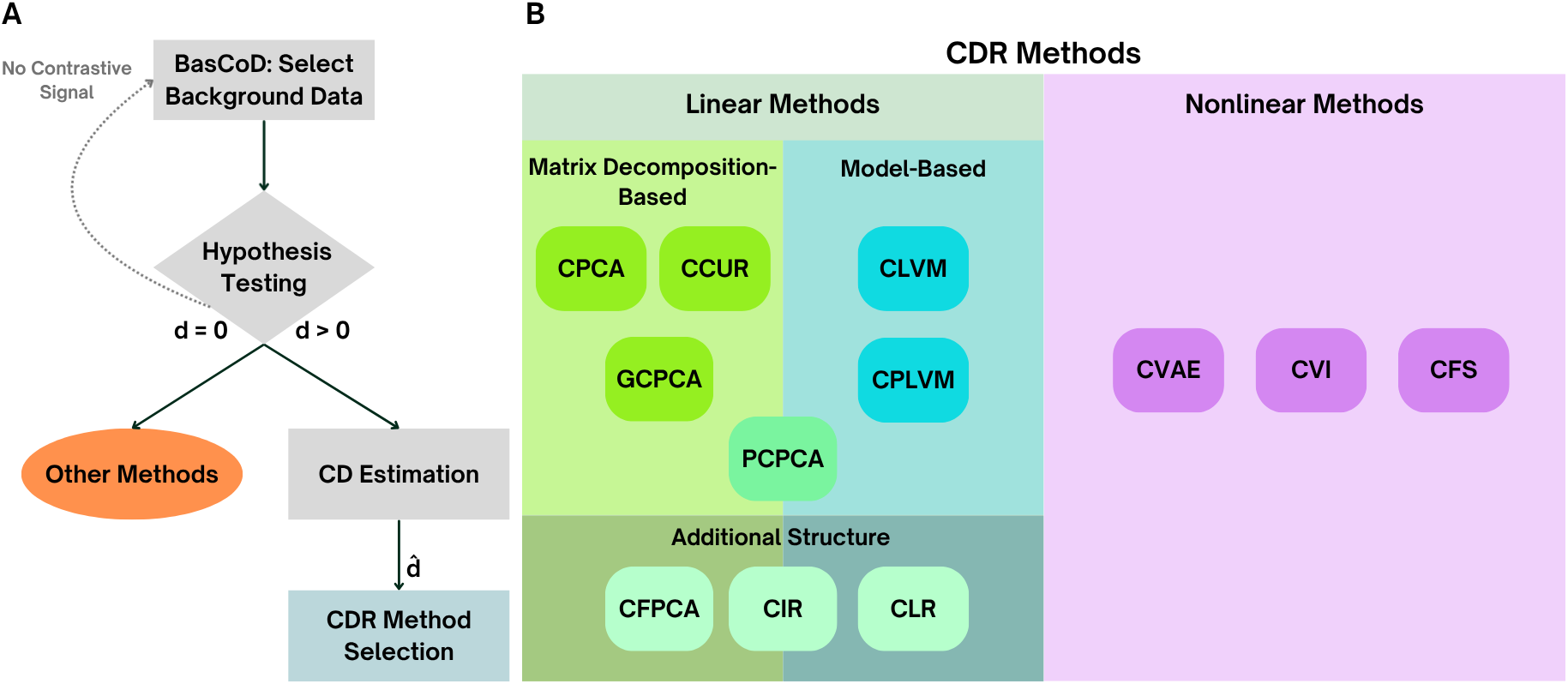}
    \caption{\textbf{Overview of CDR workflow and methods}
  \textbf{(A) Workflow.} First select an appropriate background dataset, then test for the presence of unique signal in the foreground. If no signal is detected ($d=0$), proceed with non-contrastive analyses or revisit the background choice. If a signal is present ($d>0$), estimate the contrastive dimension $\hat d$ and then choose and implement a CDR method using $\hat{d}$.
  \textbf{(B) Method taxonomy.} Representative CDR methods are organized by family with subgroups within each color-coded family. }
    \label{fig:1}
\end{figure}

When $H_0$ is rejected, CDE estimates the contrastive dimension $d$, the dimension of low-dimensional signal unique to the foreground, via thresholding the singular values $\lambda_k$. For a tolerance \(\varepsilon\in(0,1)\) chosen by the user, the estimated
\[
\widehat d\;\coloneqq\;\#\{k:\,\widehat\lambda_k<1-\varepsilon\}\;+\;\max(d_x-d_y,\,0),
\]
i.e., count principal angles exceeding \(\arccos(1-\varepsilon)\) and add the unavoidable dimension \(d_x-d_y\) when $d_x>d_y$, which automatically implies some unique information in $X$. 
Under sub-Gaussian assumptions, \(\widehat d\) is consistent with finite-sample error controlled by eigengaps and the sampling covariance matrices of $X$ and $Y$.

CDE serves as a diagnostic for whether CDR is appropriate and, if so, how to choose the contrastive dimension, a key tuning parameter in almost all CDR methods. Together with background selection, CDE structures the decision-making process summarized in \Cref{fig:1}A, while \Cref{fig:1}B shows a taxonomy of CDR methods.

\section{Experiments}

\subsection{A toy example: Corrupted MNIST}

\begin{figure}[!h]
\centering
\renewcommand{\arraystretch}{1.0} 
\setlength{\tabcolsep}{6pt}       
\begin{tabular}{ccc}
\includegraphics[width=0.25\linewidth]{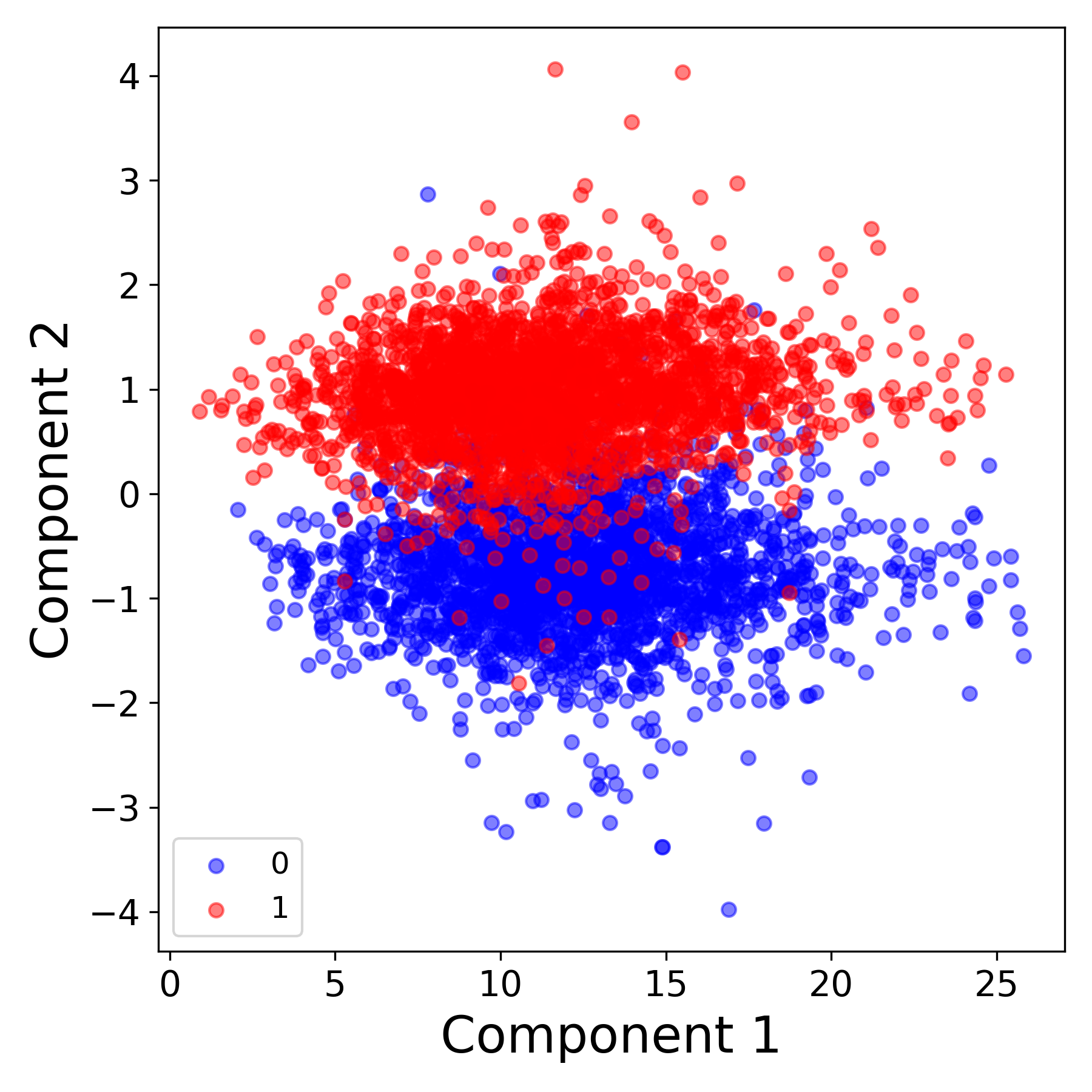} &
\includegraphics[width=0.25\linewidth]{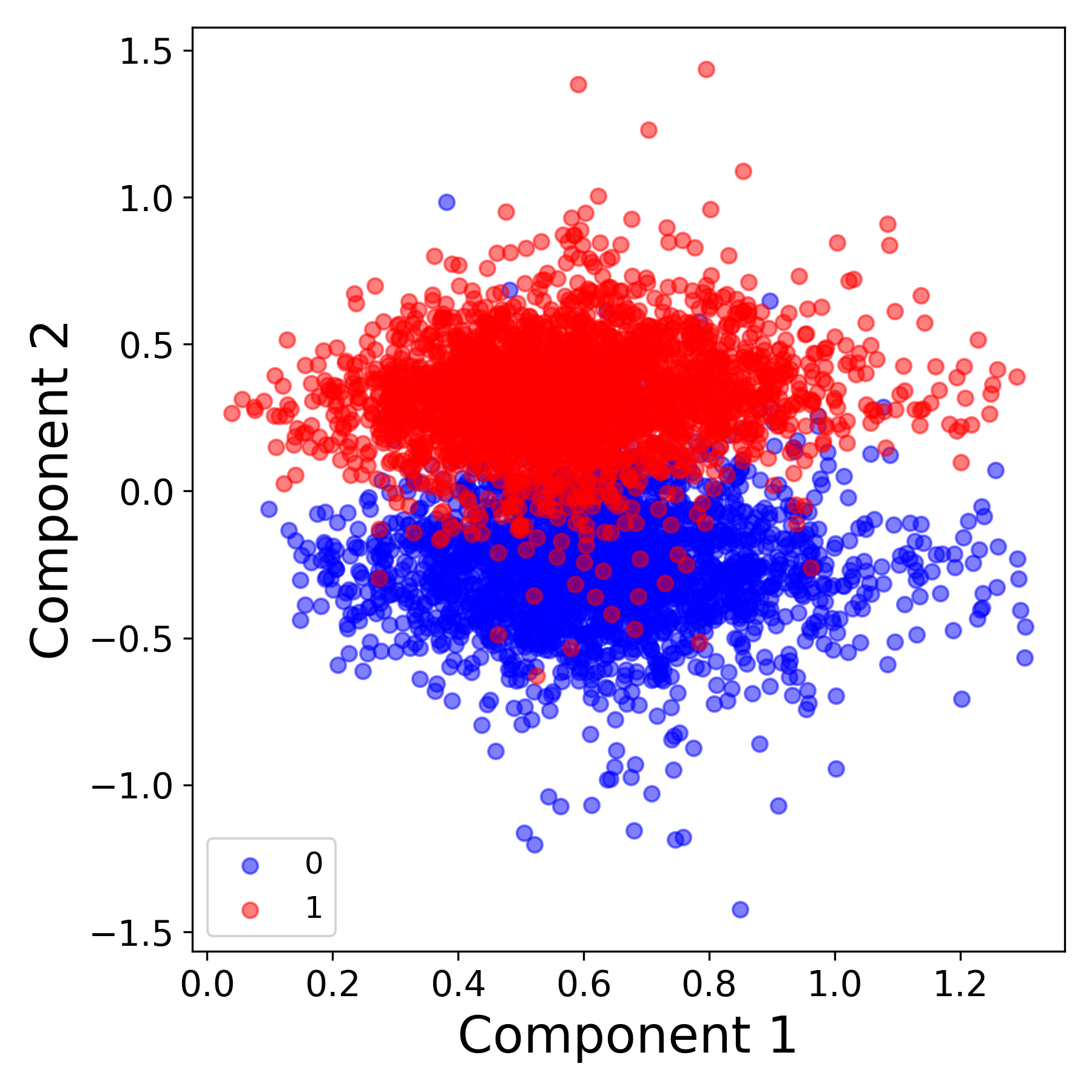} &
\includegraphics[width=0.25\linewidth]{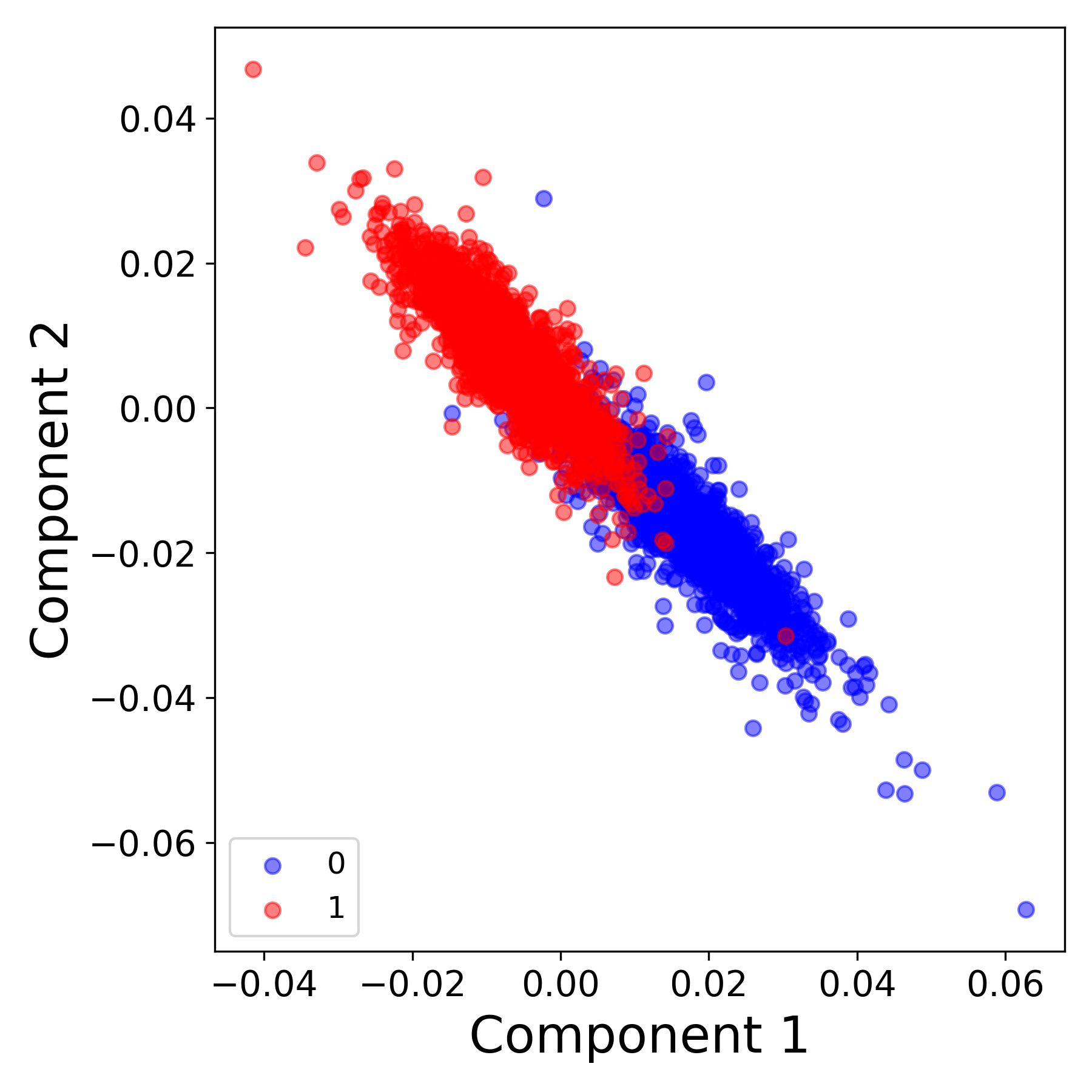} \\
{\small CPCA} & {\small PCPCA} & {\small GCPCA} \\
\includegraphics[width=0.25\linewidth]{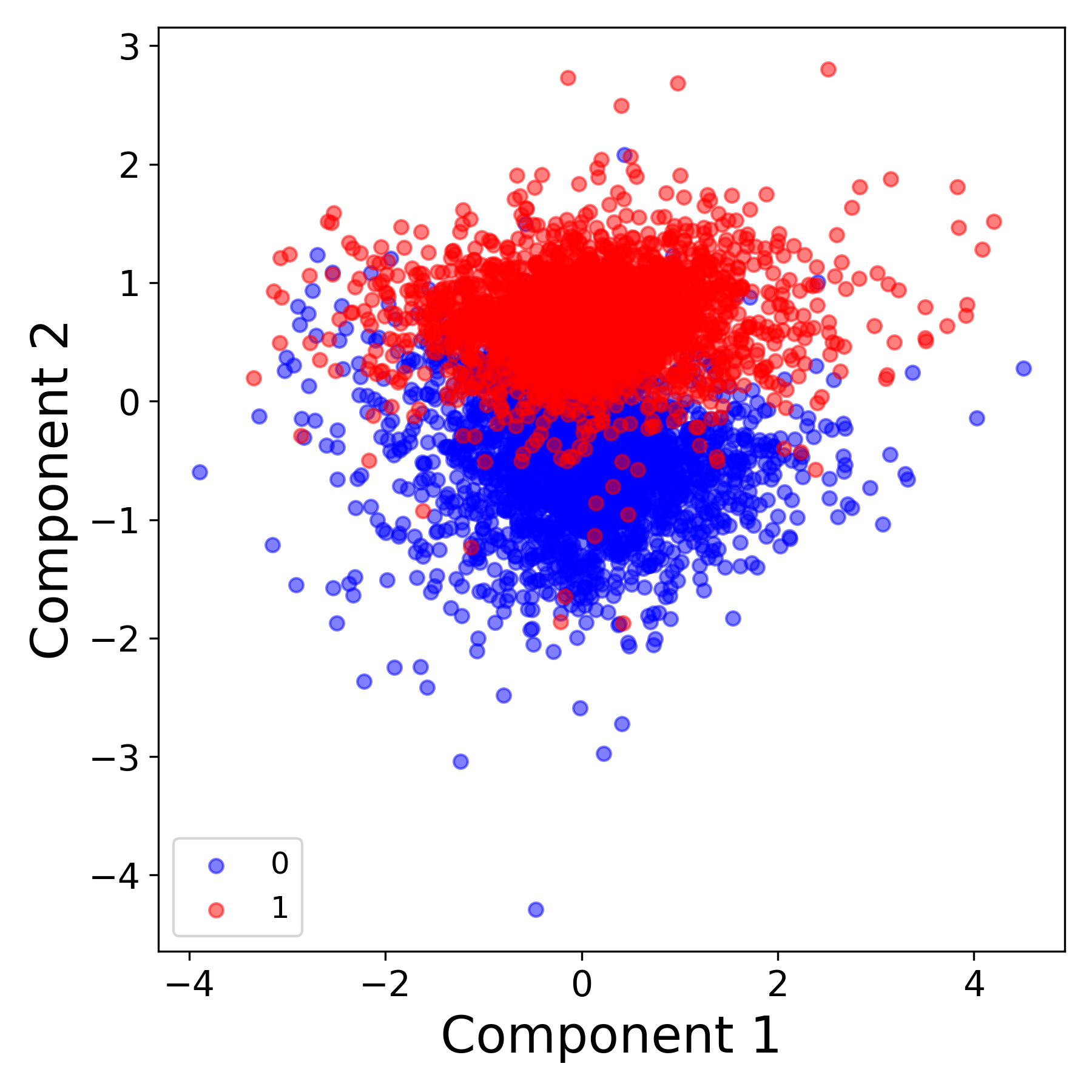} &
\includegraphics[width=0.25\linewidth]{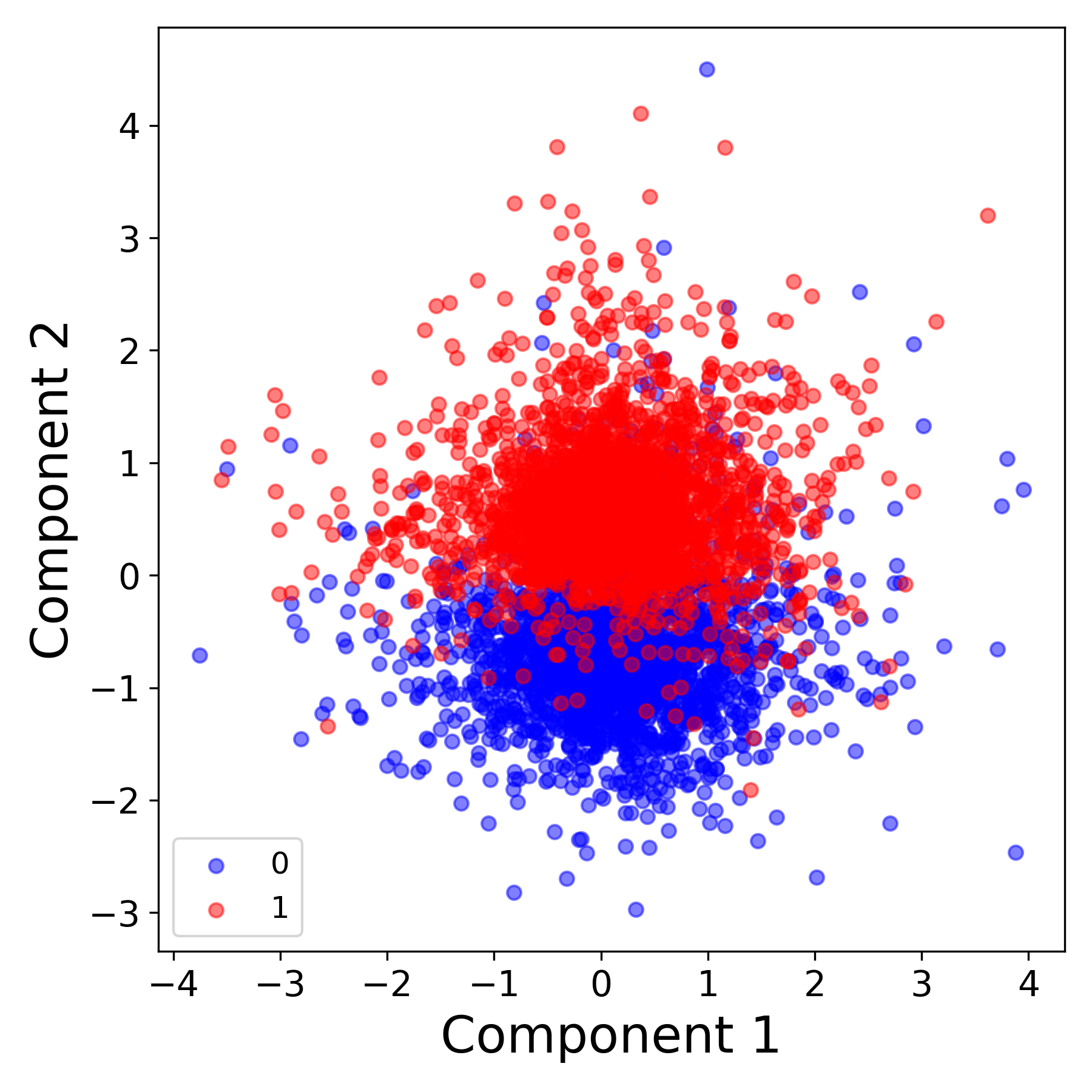} &
\includegraphics[width=0.25\linewidth]{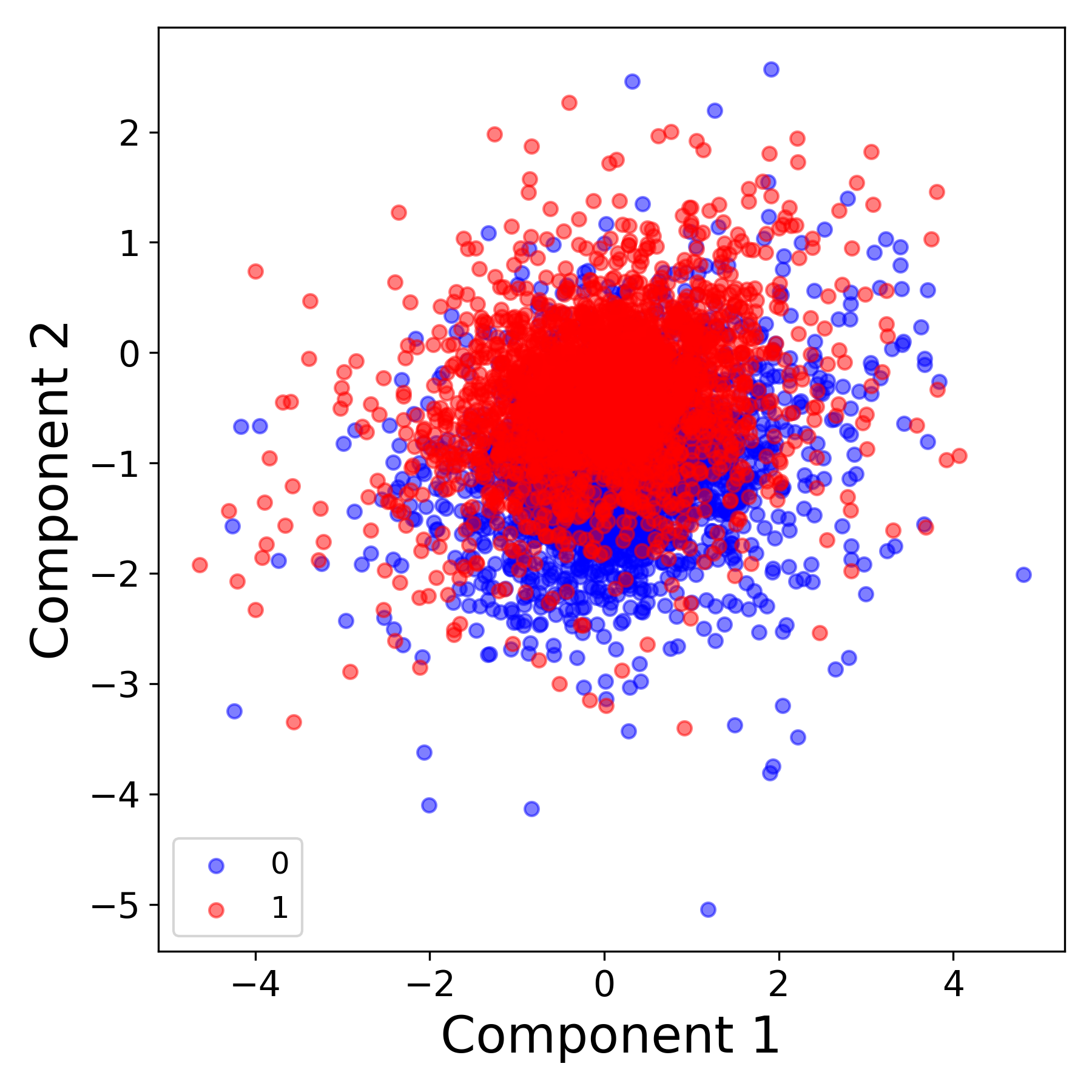} \\
{\small CVAE} & {\small CVI} & {\small CLVM} \\
\end{tabular}
\caption{Two-dimensional representation from six representative CDR methods on corrupted MNIST dataset.}
\label{fig:res_mnist}
\end{figure}

In this section, we evaluate several representative CDR methods on a synthetic dataset constructed by overlaying MNIST digits \citep{lecun1998convolutional} onto natural image backgrounds of grass (\Cref{fig:mnist_ex}). Following the setup in \citep{abid2018exploring}, a target dataset of 5000 images is created by randomly superimposing handwritten digits 0 and 1 from the MNIST dataset onto natural background textures of grass taken from the ImageNet dataset \citep{russakovsky2015imagenet}. The grass images are first converted to grayscale, resized to 100x100 pixels, and then randomly cropped to 28x28 to match the MNIST digits before overlaying. This design produces images that combine a structured foreground signal (the digits) with high-variance background noise (the grass texture), creating a useful benchmark for assessing CDR methods. The resulting 2-dimensional representations are shown in \Cref{fig:res_mnist}, where all methods achieve a reasonable degree of separation, successfully distinguishing images containing digits 0 and 1. This demonstrates their ability to extract meaningful foreground structure by leveraging a background dataset to denoise the unwanted variation.

\subsection{A case study: Mouse protein}

In this section, we evaluate representative methods on the mouse protein dataset \citep{higuera2015self}, a widely used benchmark for CDR. The dataset contains measurements of 77 protein expression levels from mice subjected to a learning experiment. The foreground group consists of 270 mice that underwent shock therapy, including both Down Syndrome (DS) and non-DS mice, while the background group contains 135 control mice without DS that did not receive shock therapy. The study was designed to investigate how exposure to shock therapy influences cognitive function, with particular interest in whether the response differs between DS and non-DS mice. 

\begin{figure}[!h]
\centering

\imgcell{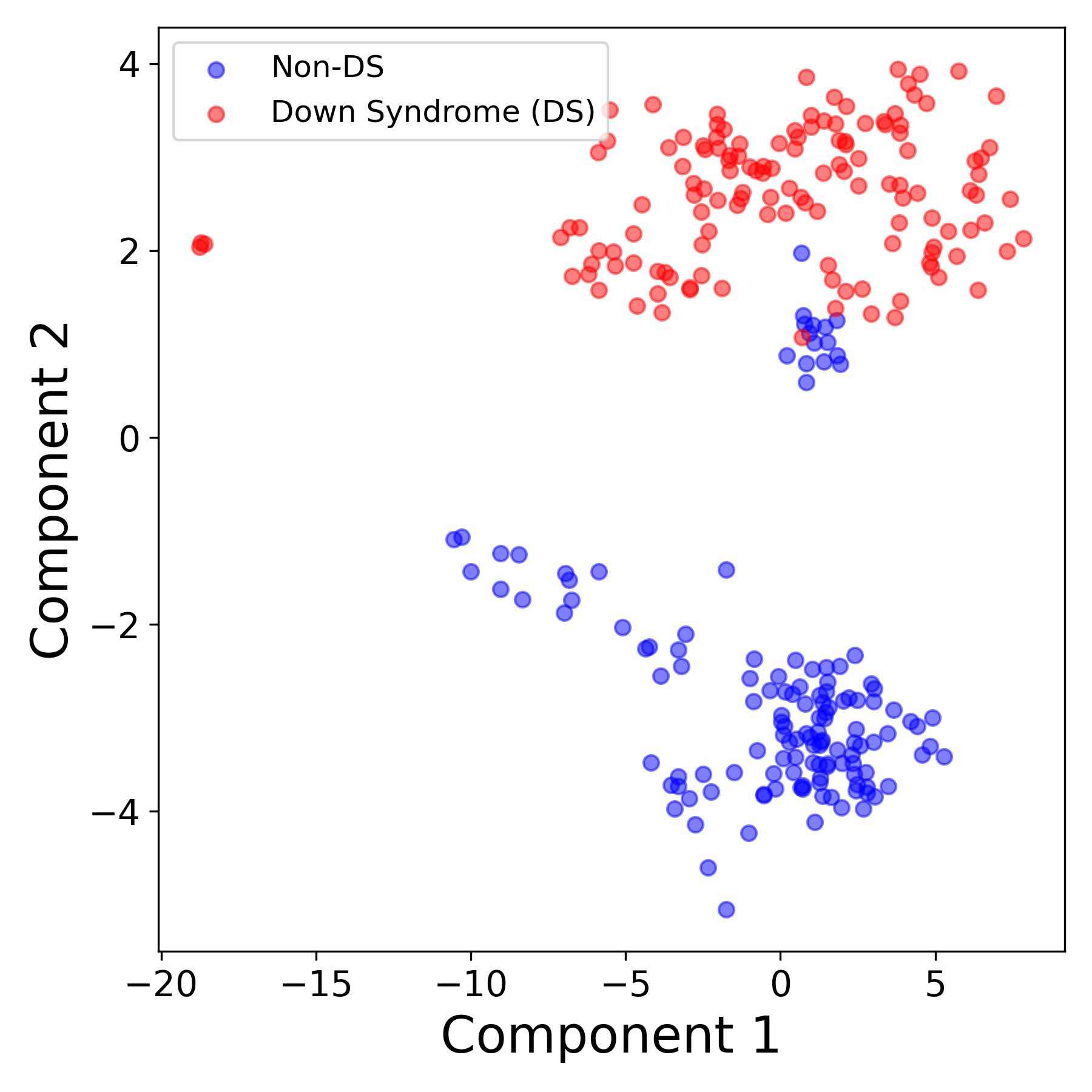}{CPCA}\hspace{0.001\linewidth}%
\imgcell{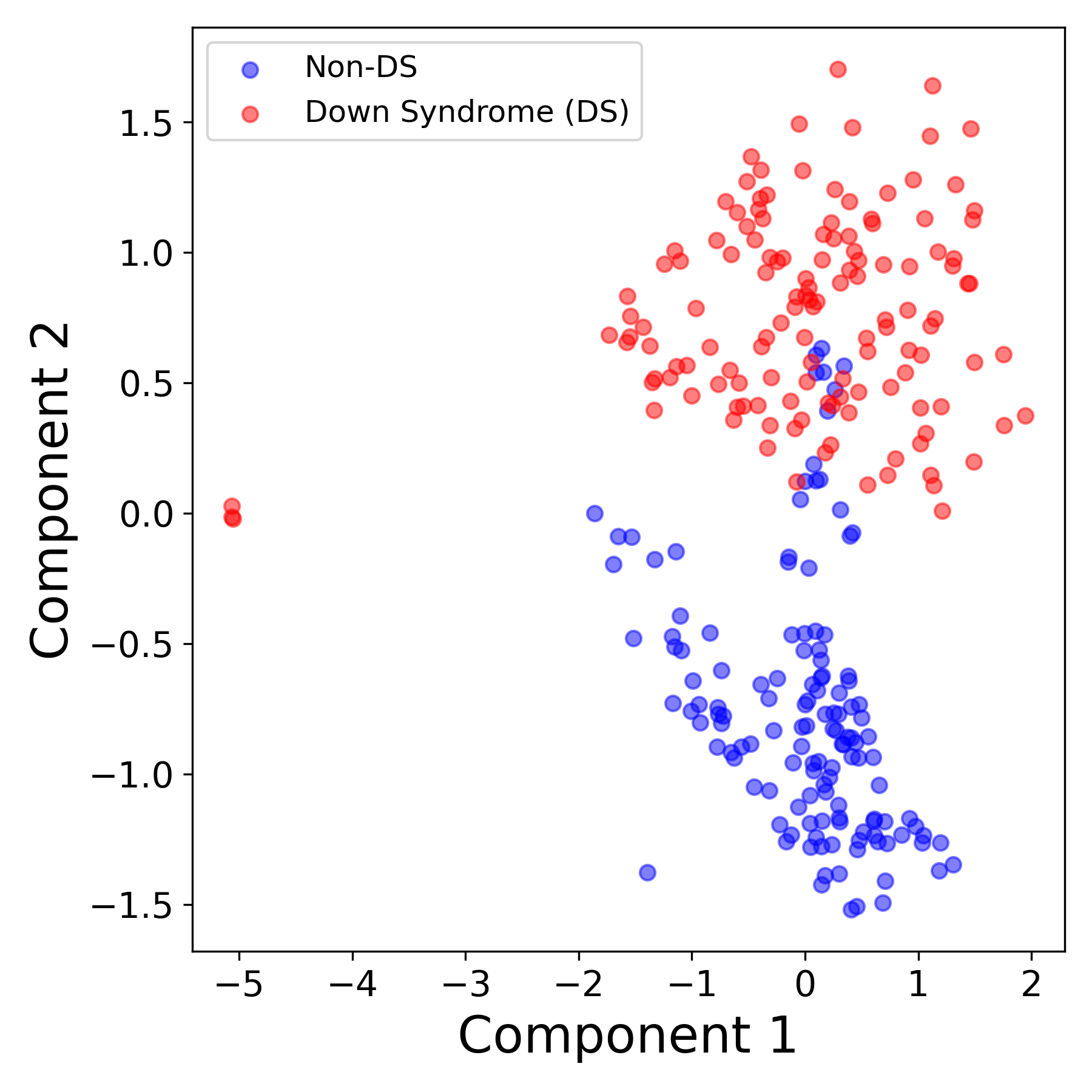}{PCPCA}\hspace{0.001\linewidth}%
\imgcell{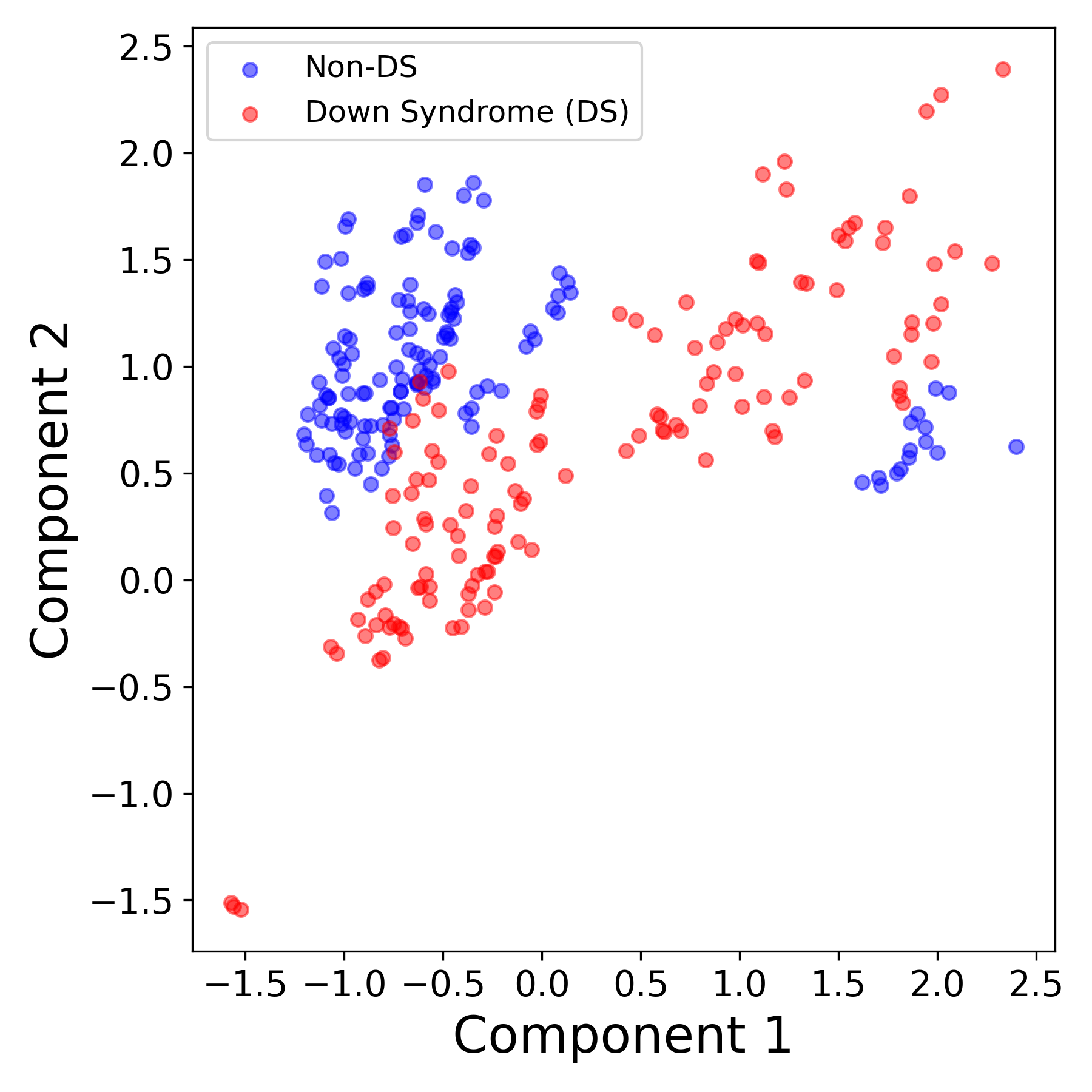}{CLVM}

\vspace{0.6em} 

\imgcell{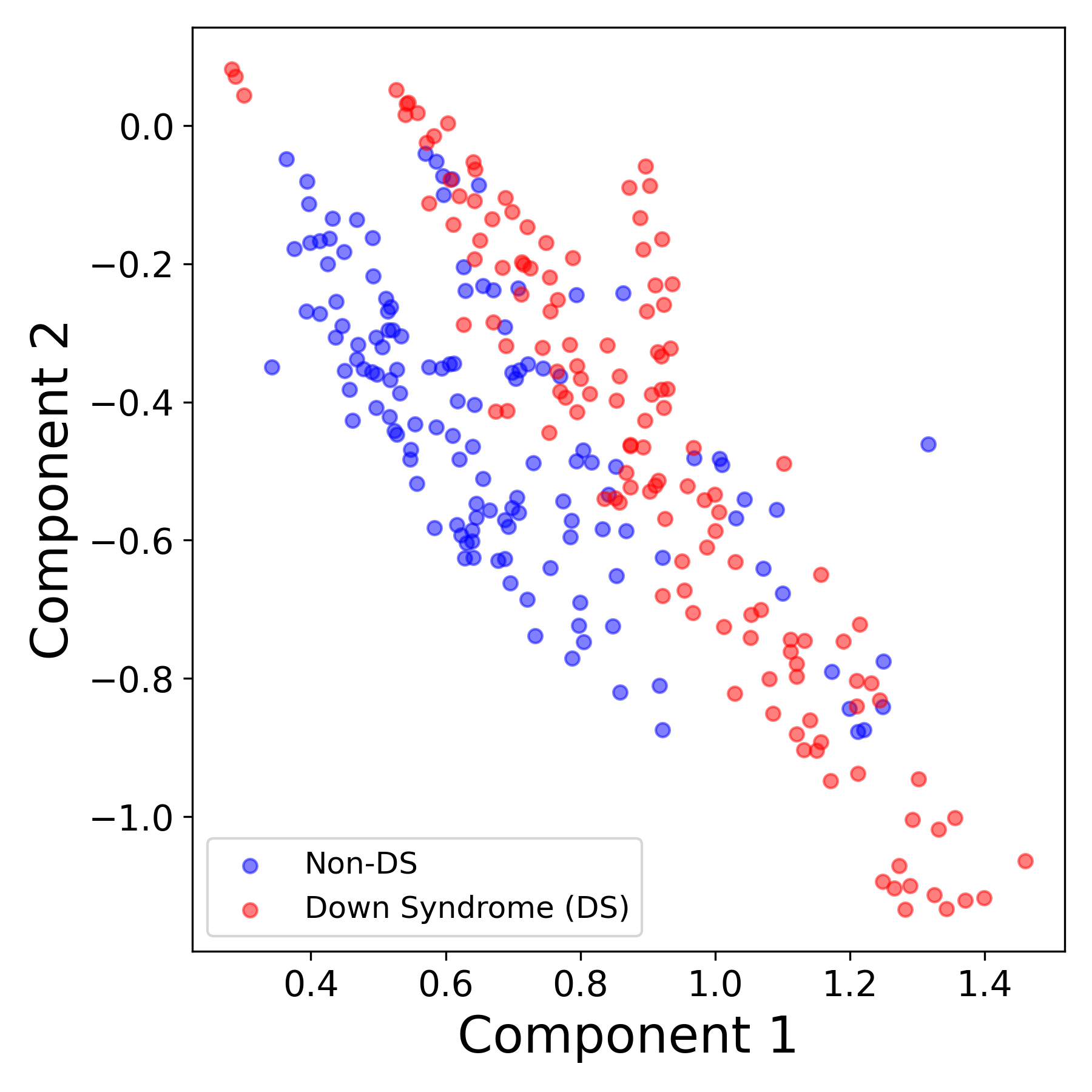}{CVAE}\hspace{0.001\linewidth}%
\imgcell{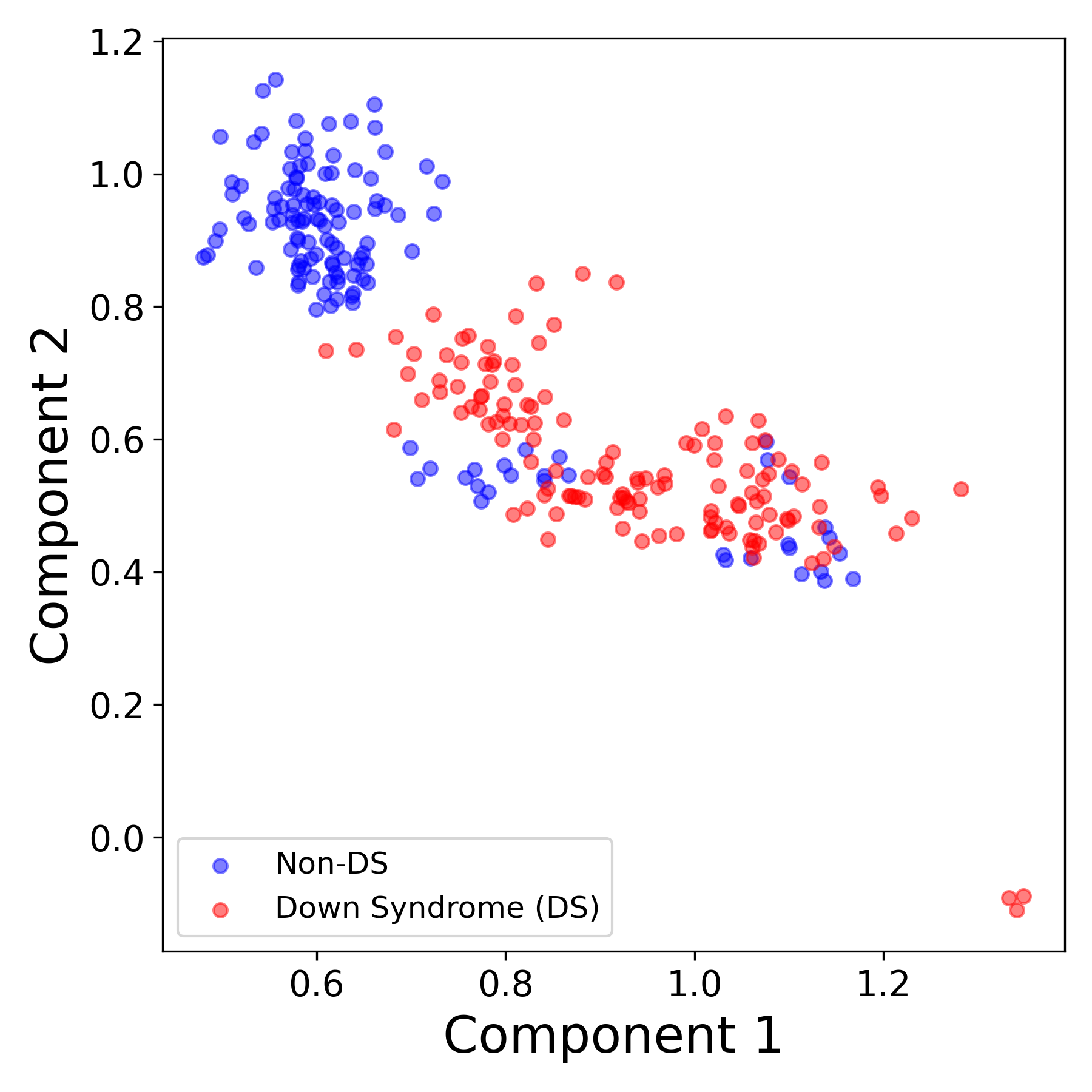}{CVI}\hspace{0.001\linewidth}%
\imgcell{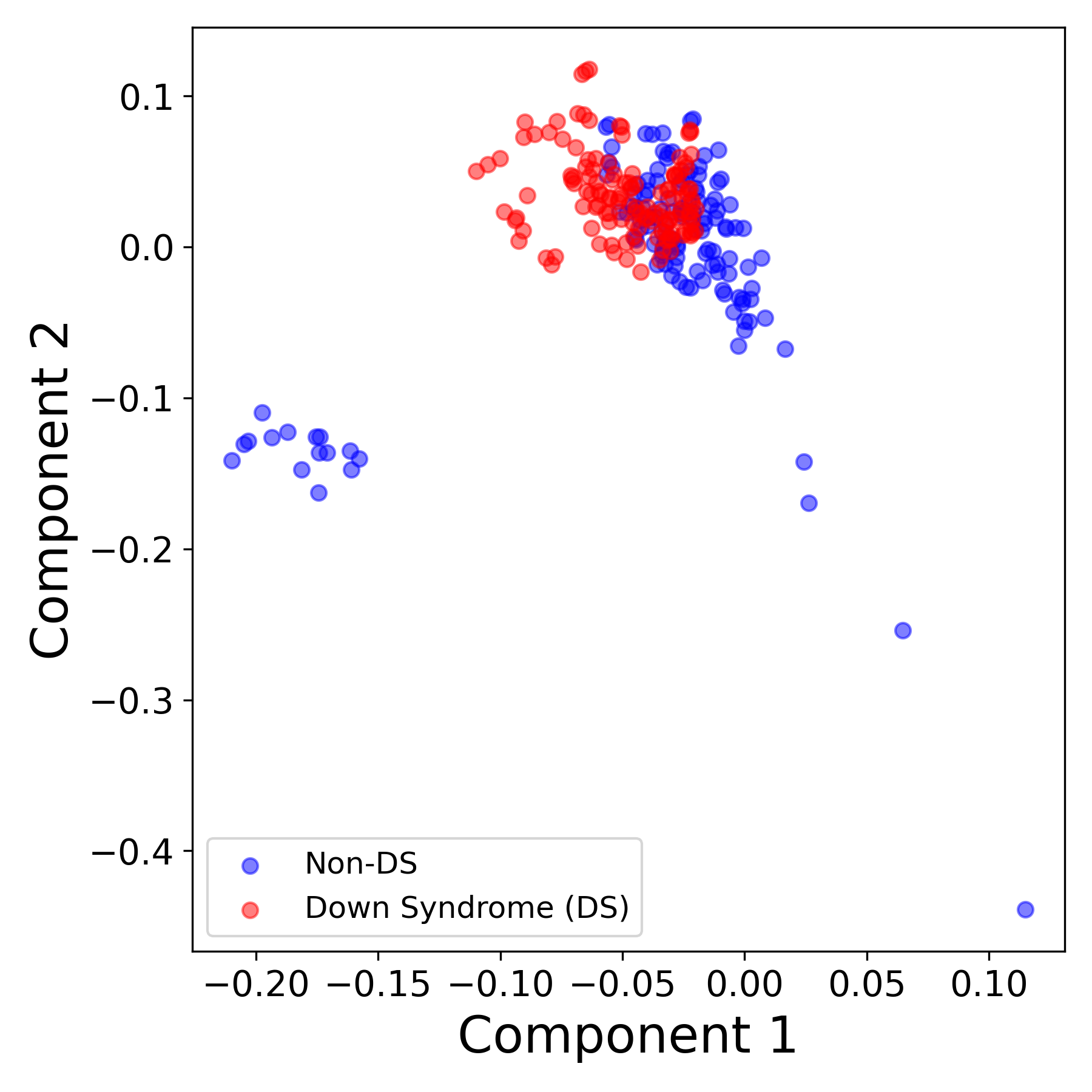}{GCPCA}

\vspace{0.6em}

\imgcell{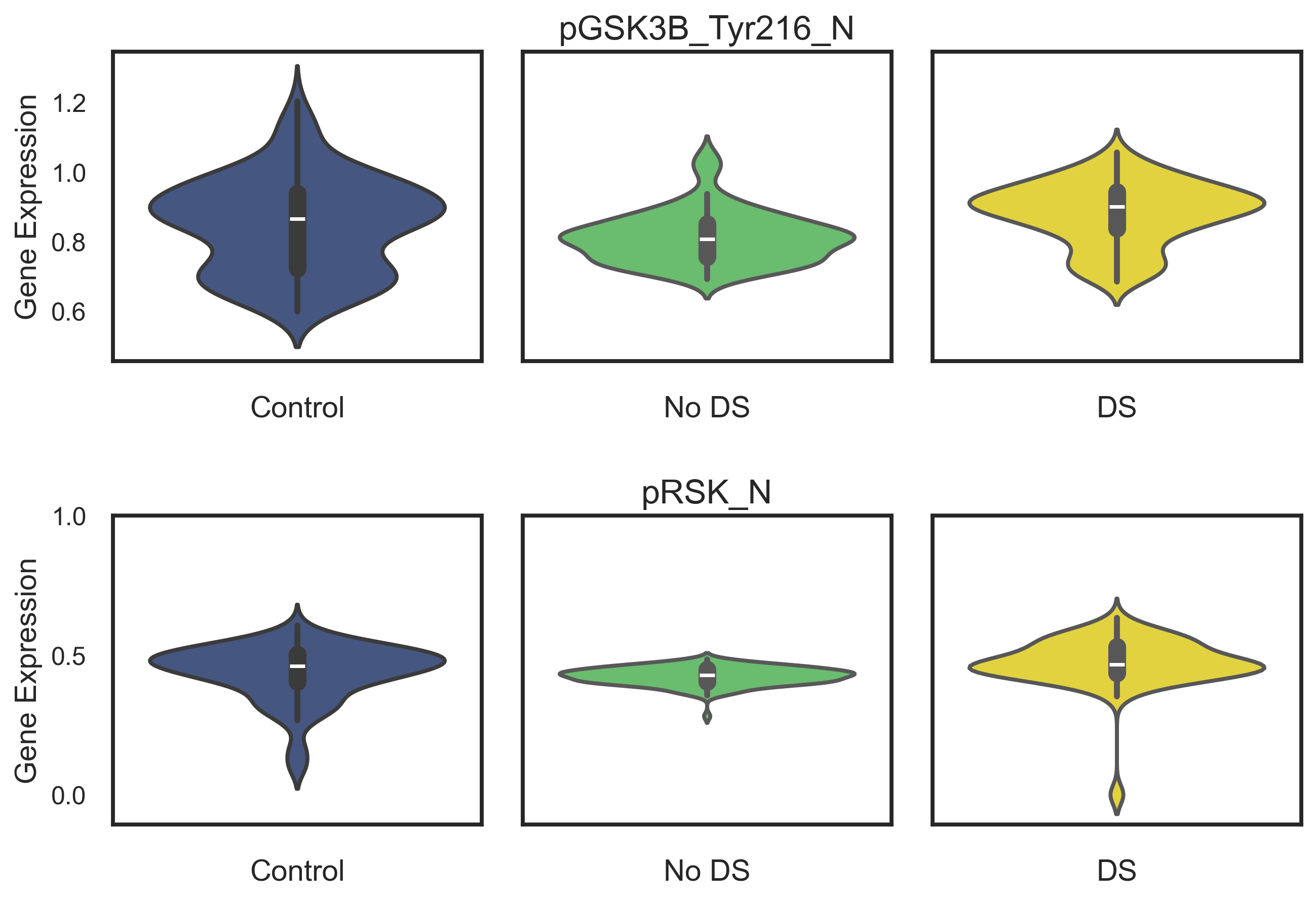}{CFS}\hspace{0.01\linewidth}%
\imgcell{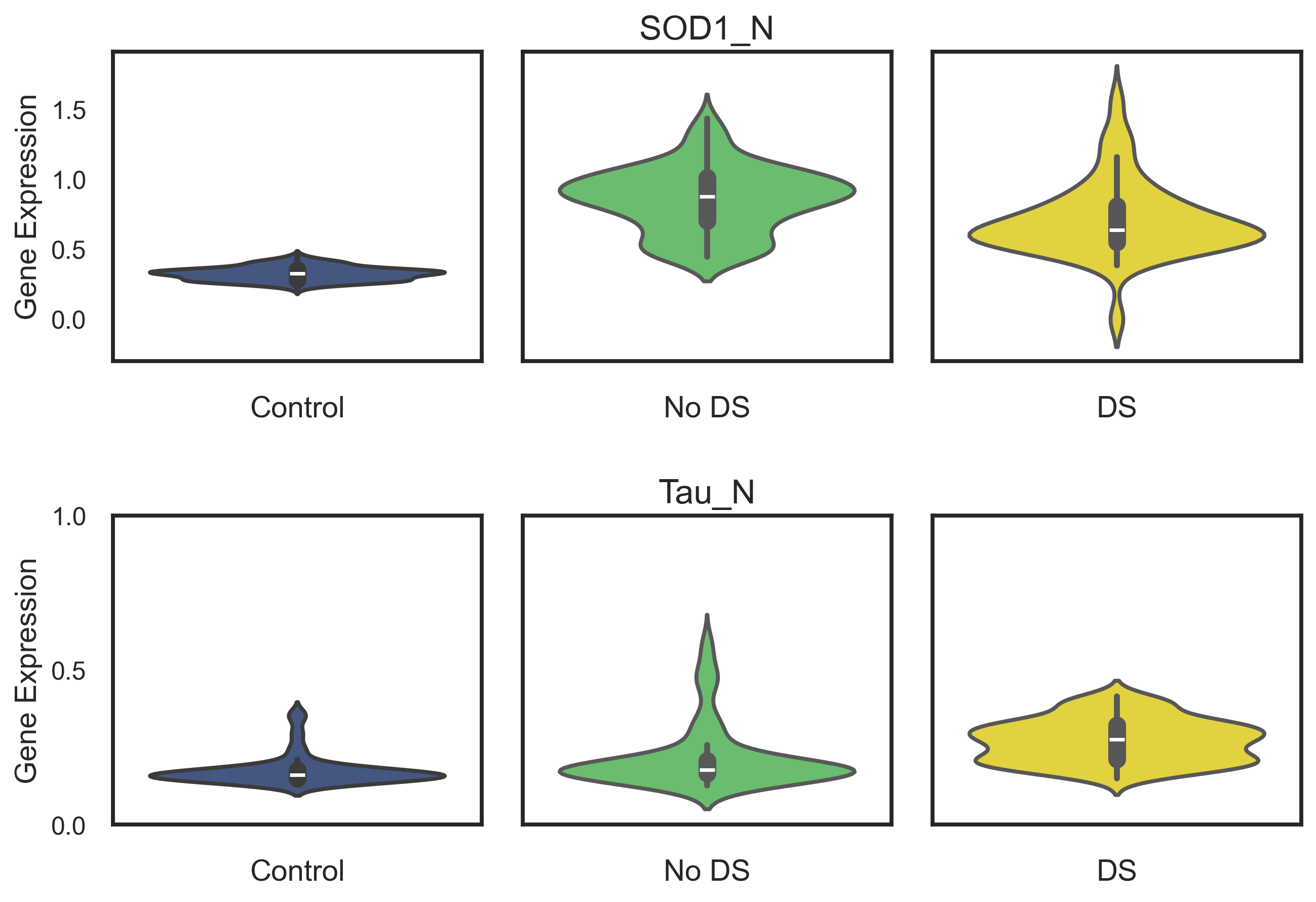}{CCUR (Columns)}\hspace{0.001\linewidth}%
\imgcell{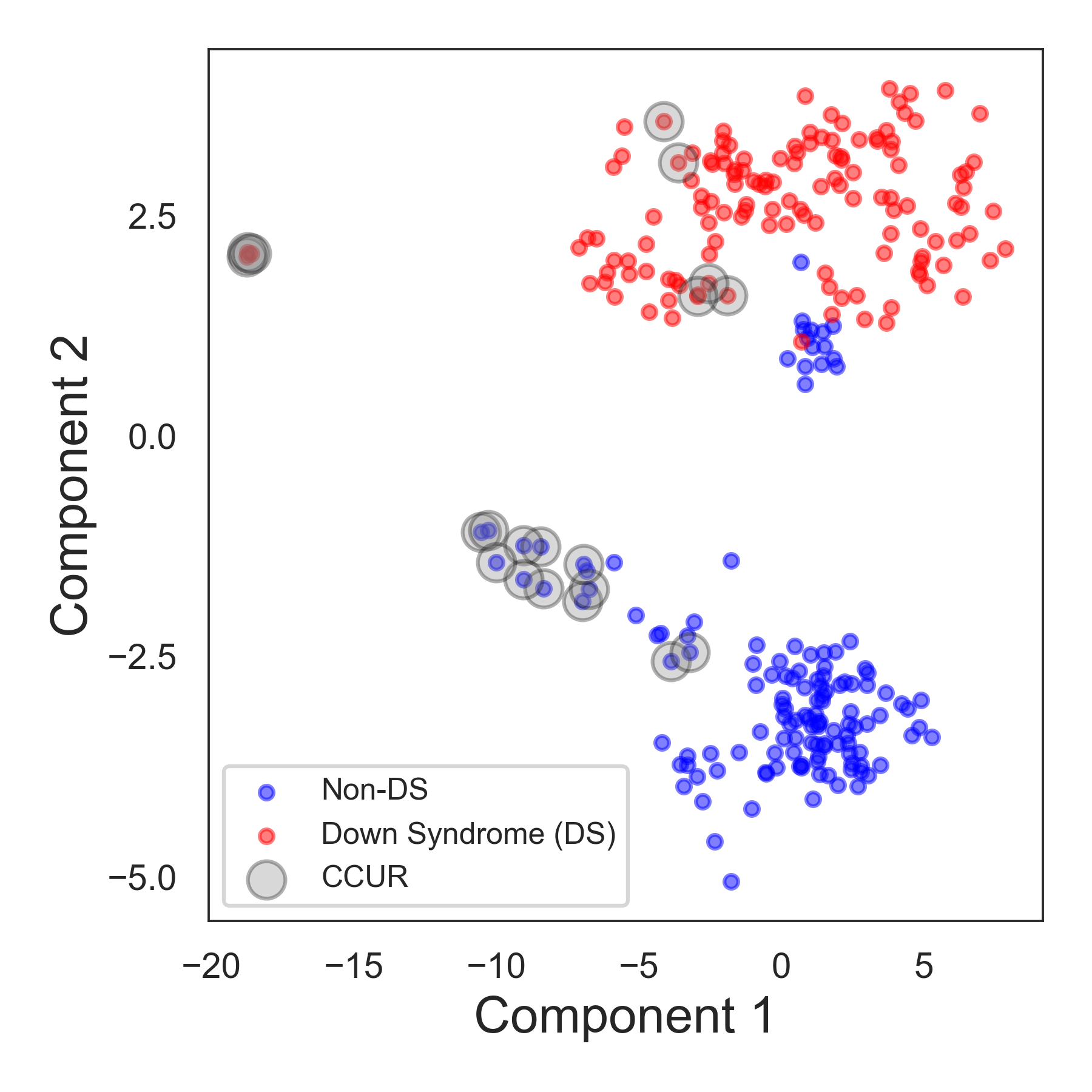}{CCUR (Rows)}

\vspace{0.6em}
\caption{Results from CDR methods on mouse protein dataset.}
\label{fig:contrastive_comparison}
\end{figure}

In the scatterplots of \Cref{fig:contrastive_comparison}, the 2-dimensional representations obtained by the CDR methods uncover well-defined DS and non-DS subgroups within the foreground that would otherwise be obscured using a non-contrastive approach. 
These results suggest that the mechanism by which shock therapy affects mice and their cognitive functions differs for mice with DS and without DS. 

In addition to visualization, we investigate feature selection results from CFS and CCUR. The violin plots reveal that both approaches not only identify proteins that distinguish between the foreground and background, but also capture notable differences within the foreground itself between mice with and without DS. For contrastive row (sample) selection, we display the 2-dimensional CPCA representations while highlighting CCUR-selected samples, showing a balanced mix of mice with DS and without DS. This is crucial for downstream analysis as the objective of this experiment was to understand how shock therapy affected mice with DS.



\section{Limitations and Future Work}


CDR has emerged as a powerful tool to analyze case-control studies with growing popularity across diverse scientific domains. Recent methodological advances have demonstrated its potential for isolating meaningful signal by leveraging appropriate background data. Despite these developments, important limitations remain, which also present opportunities for future research. In this section, we outline several promising directions to further advance the field.

\subsection{Hyperparameter Selection}



A recurring challenge across CDR methods is the reliance on hyperparameters whose influence on results is not fully understood. In most of the papers introducing these methods, guidance on hyperparameter choice is either minimal, often limited to a simple grid search, or absent altogether. While authors typically demonstrate that for some choice of hyperparameters, their proposed method can outperform existing baselines, the rationale for those choices is rarely transparent. This creates a practical barrier for domain scientists, since the burden of tuning is shifted to the practitioner without clear heuristics or theoretical guarantees.

The lack of principled hyperparameter selection raises several concerns. First, it undermines reproducibility, since different practitioners analyzing similar data may arrive at divergent results due solely to tuning choices. Second, it complicates interpretation, as it becomes unclear whether observed patterns arise from the data or from arbitrary parameter settings. Third, the computational cost of exhaustive search can be prohibitive for large-scale or high-dimensional datasets, which limits accessibility.

Therefore, an important avenue for future work is the development of fast, data-driven, and goal-oriented approaches to hyperparameter selection. Possible directions include the use of stability-based criteria, cross-validation schemes adapted to the contrastive setting, Bayesian optimization strategies, and information-theoretic measures that connect tuning parameters to identifiable structure in the data. In addition, theoretical work is needed to characterize the sensitivity of methods to hyperparameters and to provide principled defaults that balance generality and interpretability. Addressing these issues would not only enhance the usability of CDR methods but also increase their reliability and adoption in scientific domains.

\subsection{Interpretability}




A key issue that may deter practitioners from utilizing CDR methods is the difficulty of interpretation. While dimension reduction methods in general are motivated by the need to compress a dataset without losing too much information, the motivation of CDR is specifically to isolate the signal unique to one group. In this setting, interpretability becomes especially important. If researchers cannot connect the reduced representation back to meaningful scientific variables, the method is unlikely to see widespread adoption.

Consider the case of a gene expression study comparing treatment and control groups. A method such as CPCA may identify a factor loading $V$ that captures high variation in the treatment group and low variation in the control group. Yet interpreting $V$ itself is not straightforward. A heatmap of its entries, with rows labeled by gene names, can offer a descriptive visualization, but it does not provide a principled explanation of which genes or biological pathways are most influential. This gap between identifying structure and explaining it highlights a central obstacle to the practical use of CDR. While interpretability was a large part of the motivation behind CCUR \citep{zhang2025contrastive1} and CFS \citep{weinberger2023feature}, there are additional future directions to pursue in this area.

One promising such direction is the development of sparse CDR methods. By incorporating penalties such as the $L_1$ norm into the optimization problems described in \cref{subsection:linear}, it may be possible to obtain factor loadings with only a small number of nonzero entries. Such sparsity would directly identify the features most responsible for group-specific variation, offering practitioners a clearer scientific story. Beyond simple sparsity, structured regularization could encourage groups of features (such as sets of genes in the same pathway) or enforce hierarchical interpretability.

Future work could also explore strategies for interpretability that go beyond sparsity. Rotations of the learned subspace, post-hoc feature scoring, and connections to variable importance measures may all help bridge the gap between statistical representation and domain knowledge. In parallel, theoretical work is needed to formalize the trade-off between interpretability and performance: sparse solutions may sacrifice subtle but meaningful patterns, while dense solutions may obscure the main drivers of variation. A systematic study of these trade-offs would provide much-needed guidance to practitioners.

Developing interpretable and sparse methods has the potential to increase the popularity of CDR approaches in applied research. More importantly, it would align these methods with the central motivation of CDR: not only to detect signal that is unique to one group, but also to communicate clearly what that signal represents.

\subsection{CDE in Nonlinear Setting}

Even in the linear setting, estimating the appropriate reduced dimension $d$ for CDR presents substantial challenges. Existing work has proposed approaches that rely on separate estimates of the intrinsic dimension of the foreground and background datasets \citep{hawke2024contrastive}. While such methods provide a starting point, they inherit the difficulties of intrinsic dimension estimation itself, which is highly sensitive to methodological choices. As a result, even in linear CDR, choosing $d$ in a principled and reproducible way remains an open problem.

While these issues are already substantial in the linear setting, they become even more pronounced when the data are believed to lie on curved manifolds, which is arguably the more general case in many applications. In linear CDR, the contrastive dimension can be understood in terms of subspaces $V_X$ and $V_Y$ estimated from the foreground and background. Extending this idea, one might instead seek to define contrastive dimension in terms of differences in geometry between manifolds $M_X$ and $M_Y$. One key question is how to formalize such a notion: should it be the minimal number of degrees of freedom needed to describe the variation that separates $M_X$ from $M_Y$, or another measure of the additional complexity in the foreground relative to the background?

In practice, progress on this problem would have immediate implications for the usability of CDR methods. A reliable notion of contrastive dimension in the nonlinear setting could directly inform the choice of the reduced dimension parameter $d$, which is currently left to ad hoc heuristics or computationally expensive tuning. By grounding $d$ in the underlying geometry of the foreground and background, researchers could obtain representations that are both more principled and more reproducible. This makes nonlinear contrastive dimension estimation not only a theoretical challenge but also a practical priority for advancing CDR methods.



\subsection{Multiple and Continuous Treatment Settings}



The methods highlighted in this review all require datasets to be partitioned into two groups: a foreground and a background. A natural question is whether there is a clear way to extend these methods to datasets with three or more groups, such as multiple treatments and multiple control groups. This generalization is highly relevant in practice, since many scientific studies are designed with several experimental conditions, disease subtypes, or longitudinal stages that cannot be adequately captured by a simple two-group comparison. However, a straightforward extension is not obvious, because the way that additional groups should influence the reduced representation is not well defined.

To formalize one version of the problem, suppose we observe foreground datasets $X_1 \in \mathbb{R}^{n_1 \times p}$ and $X_2 \in \mathbb{R}^{n_2 \times p}$ along with a background dataset $Y \in \mathbb{R}^{m \times p}$. We may not expect $X_1$ and $X_2$ to arise from the same distribution, yet we may wish to find a factor loading $V \in \mathbb{R}^{p \times d}$ that isolates the information unique to $X_1$ relative to $Y$, while also leveraging the information contained in $X_2$. 

An equally important but distinct challenge arises in the case of continuous treatments, for example varying drug dosages, developmental time courses, or disease progression stages. In these scenarios, the contrast is not defined by sharp boundaries between groups, but rather by gradual changes in the data distribution along a continuum. This illustrates the broader challenge: how should additional or continuous treatments and backgrounds be incorporated into contrastive objectives in a way that is both principled and interpretable?

Future work could explore several directions. One is to design multi-objective formulations where each treatment-control comparison contributes a separate contrastive objective, and the resulting representation balances these objectives in a principled way. Another is to develop models that explicitly capture shared versus group-specific structure, for instance through hierarchical decompositions or tensor factorizations. A third direction is to clarify what should count as “signal unique to one group” when multiple groups overlap in complex ways. For example, a pattern that is present in two treatment groups but absent in controls may or may not be considered unique to each treatment, depending on the definition. In parallel, extending contrastive methods to continuous treatments, such as drug dosage or time-course experiments, represents another promising avenue, where the goal would be to extract low-dimensional structure that varies systematically with the continuous treatment variable while filtering out background effects. 

\subsection{Uncertainty Quantification}

Most existing CDR methods focus solely on extracting low-dimensional representations, with limited attention to uncertainty quantification (UQ). While a few model-based approaches, such as CLVM or PCPCA, may yield uncertainty estimates through the likelihoods, the majority of CDR  methods, particularly those relying on deep learning, do not provide calibrated uncertainty measures. This lack of uncertainty limits the interpretability and reliability of contrastive representations in downstream analyses.

Developing general-purpose UQ procedures for CDR is therefore an important direction for future work. Uncertainty estimates are essential for assessing the confidence of scientific conclusions drawn from contrastive representations and for distinguishing meaningful signal from noise. One promising avenue is to adapt model-agnostic approaches such as conformal inference~\citep{shafer2008tutorial}, which can provide finite-sample guarantees under minimal assumptions. For instance, one could construct conformal prediction sets in the contrastive embedding space or assess the stability of selected features across multiple perturbations of the background. Integrating such tools into existing CDR frameworks would enhance their robustness, increase trust in their outputs, and facilitate principled decision-making in scientific applications.

\subsection{Extensions to Multi-Modal Data}

Many modern scientific studies collect multi-modal data, such as genomics paired with imaging, behavioral measures paired with text, or electronic health records that combine structured and unstructured sources. In these contexts, both the foreground and background groups may themselves be multi-modal, and the key signals of interest may reside not only within each modality but also in their interactions. Current CDR methods are almost exclusively designed for single-modality data, which limits their applicability to these increasingly common datasets.

Extending CDR to multi-modal settings introduces several challenges. One difficulty is how to define the foreground–background comparison when signals are distributed across heterogeneous feature spaces. Should each modality be analyzed separately with its own foreground–background decomposition, or should a joint representation be constructed that captures patterns spanning multiple modalities? Another challenge is alignment: foreground and background groups may not have the same modalities observed, or may have them measured on very different scales, making it unclear how to balance their contributions. Finally, there is the question of interpretability. Even if a joint low-dimensional representation can be obtained, it must be translated back into meaningful insights within and across modalities to be useful for practitioners.

Several promising directions could be explored. Multi-view learning frameworks such as canonical correlation analysis and its extensions \citep{hardoon2004cca,andrew2013dcca,wang2015deepmv} provide natural starting points for integrating multiple modalities, since they are already designed to uncover shared and distinct structure across heterogeneous datasets. Coupled matrix and tensor factorization methods \citep{acar2011allatonce,lock2013jive} represent another promising foundation, as they explicitly model variation that is shared versus unique to each modality. Deep learning architectures have also been widely studied for aligning multi-modal data before applying downstream objectives \citep{ngiam2011multimodal,baltrusaitis2019survey}, although their use in the CDR context would raise important questions of stability and interpretability. Finally, there is an opportunity to develop modality-specific decompositions that are later integrated into a unified representation, which may offer a flexible compromise between within-modality clarity and cross-modality integration.

Developing multi-modal CDR methods would substantially broaden the scope of the field, making it relevant to emerging applications in neuroscience, biomedicine, and the social sciences where multi-modal data are now the norm. 

\subsection{Relation to Contrastive Learning}

Contrastive learning has recently emerged as a powerful tool in machine learning and artificial intelligence, where the central idea is to learn representations by comparing positive and negative pairs of data. By encouraging similar samples (positive samples) to be embedded closer and have similar representations, while pushing dissimilar samples (negative pairs) apart, contrastive learning has proven highly effective for self-supervised representation learning across domains such as computer vision~\citep{he2020momentum,chen2020simple}, language~\citep{gao2021simcse}, and biology~\citep{li2025sccontrast}. Beyond single-modality settings, contrastive learning also extends naturally to multi-modal data; for instance, paired image–text data with a contrastive loss enables the learning of shared representations across modalities, as demonstrated by the CLIP model~\citep{radford2021learning}.

Instead, we focus on CDR in this article, which emphasizes comparisons between target and background datasets. Both frameworks emphasize learning from relative comparisons rather than absolute measurements, but their contrast structures differ substantially: contrastive learning builds synthetic pairs through augmentation, while CDR leverages the natural contrast between foreground and background datasets. However, the similarity in terminology has caused confusion among some practitioners, especially those new to the field, who may conflate contrastive learning with CDR.

As a result, establishing a formal connection between them is a promising direction for future research. Contrastive learning has seen rapid algorithmic progress, including the development of novel loss functions, augmentation strategies, and sampling schemes. Adapting these innovations to CDR may improve its performance, scalability, and robustness—particularly in high-dimensional or multi-modal settings. Conversely, the structured foreground–background framework in CDR offers a principled approach to disentangling relevant signal from nuisance variation, which could enhance interpretability in contrastive learning models.

A unified perspective may also clarify the theoretical foundations of both fields, enabling a better understanding of what types of contrastive structures yield meaningful representations. For example, identifying conditions under which synthetic contrast (from augmentations) approximates natural contrast (from foreground/background separation) could inform model design across domains. More broadly, bridging the two areas would promote methodological coherence, reduce confusion among practitioners, and accelerate the development of contrastive techniques applicable to a wider range of scientific problems.
 






\end{document}